\documentclass[]{article}
\usepackage{emulateapj}
\usepackage{psfig}

\advance \voffset by 0.00cm\relax

\begin{document}

\title{STUDY OF GAMMA RAY BURST BINARY PROGENITORS}

 \author{KRZYSZTOF BELCZYNSKI\altaffilmark{1,2,3}, 
         TOMASZ BULIK\altaffilmark{3}
         AND BRONISLAW RUDAK\altaffilmark{4}}

 \affil{
     $^{1}$ Northwestern University, Dept. of Physics 
     \& Astronomy, 2145 Sheridan Rd., \#F325, Evanston, IL 60208, USA;\\
     $^{2}$Harvard--Smithsonian Center for Astrophysics, 
     60 Garden St., Cambridge, MA 02138;\\
     $^{3}$ Nicolaus Copernicus Astronomical Center,
     Bartycka 18, 00-716 Warszawa, Poland;\\
     $^{4}$ Nicolaus Copernicus Astronomical Center,
     Rabianska 8, 87-100 Torun, Poland;\\
     belczynski@northwestern.edu, bulik@camk.edu.pl, bronek@ncac.torun.pl}

\begin{abstract}
Recently much work in studying Gamma-Ray Burst (GRB) has been devoted
to revealing the nature of outburst mechanism and studies of GRB
afterglows.
These issues have also been closely followed by the quest for identifying
GRB progenitors.
Several types of progenitors have been proposed for GRBs:
the most promising objects seem to be collapsars, compact object
binaries, mergers of compact objects with helium cores of
evolved stars in common envelope episodes, and also the recently 
discussed connection of GRBs with supernovae.
In this paper we consider the binary star progenitors of GRBs:
white dwarf neutron star binaries (WD-NS), white dwarf black hole
binaries(WD-BH),
helium core neutron star mergers (He-NS), helium core black hole
mergers (He-BH), double neutron stars (NS-NS) and 
neutron star black hole binaries (NS-BH).

Using population synthesis methods we calculate merger rates of
these
binary progenitors and  we compare them to the observed BATSE GRB rate.
For  the binaries considered, we also calculate the distribution of merger
sites around  host galaxies and compare them to the observed
locations of  GRB afterglows with respect to their hosts.
We find that the rates of binary GRB progenitors in
our  standard model are lower than  the observed GRB rates
if  GRBs are highly collimated.     However, the uncertainty in the 
population synthesis results 
is too large to make this a firm conclusion.
Although some observational signatures seem to point to collapsars
as progenitors of long GRBs, we find that mergers of WD-NS,
He-NS, He-BH, and NS-NS systems also trace the star formation regions
of their host galaxies, as it is observed for long GRBs.

We also speculate about possible progenitors of short-duration GRBs.
For these, the most likely candidates are still mergers of compact
objects.
We find that the  locations NS-NS and NS-BH mergers
with respect to their hosts are significantly different.
This may allow to distinguish between these two progenitor
models, once current and near future missions, such as HETE-II or
SWIFT, measure the locations of short GRBs.
\end{abstract}


\section{INTRODUCTION}

The last decade  brought a great breakthrough in gamma-ray
burst  studies.  The BATSE detectors on GRO have shown that GRBs
are distributed isotropically on the sky, and that
their brightness distribution is not consistent with
a uniform  source distribution in Euclidean space (Paciesas et al. 1999).
Observations of GRB afterglows in X-ray, optical
and radio wavelength  domains (Costa et al.\ 1997;
Groot et al. \ 1997b) led to identification of GRB host galaxies
(Groot et al. \ 1997a) and measurements of their redshifts.
This has solved the long
standing problem of their distance scale. While we learned that GRBs
come from cosmological distances, there  are still two major
difficulties in understanding this phenomenon. First, we do not
fully understand the physics of the outburst. Although several models
have been proposed, they all have yet to  meet some severe
constraints imposed by observations (i.e. releasing  energies of
$10^{51}$--$10^{54}$\ ergs in timescales as short as
$10^{-2}\,$s
in the case of some GRBs). Second we do not know what are the
astronomical objects leading to  gamma-ray bursts, i.e. what are
their progenitors.

In recent years the black hole accretion disk model for GRBs has
been  given much attention (Fryer, Woosley \& Hartmann 1999a; Meszaros 2000; 
Brown et al.\ 2000).
Progenitors leading to this model include  collapsars
(Woosley 1993; Paczynski 1998, MacFadyen \& Woosley 1999) and binary mergers:
helium star black hole (Fryer \& Woosley 1998),
double neutron stars (Ruffert et al.\ 1997; Meszaros \& Rees 1997),
black hole neutron star (Lee \& Kluzniak 1995; Kluzniak \& Lee 1998)
and  black hole white dwarf systems (Fryer et al.\ 1999b).
Also recently  the connection between supernovae and gamma-ray bursts 
received much attention (Paczynski 1999; Woosley 2000; Chevalier 2000),
however, there is  still no clear evidence that these two phenomena are
intrinsically  correlated (Graziani et al.\ 1999).

A good method of discerning among the binary progenitors is to
compare theoretical predictions of their merger site distributions around 
host galaxies with location of observed GRBs within host galaxies.
Binary population synthesis can be of great help in addressing this
question.
One can calculate the properties of a
given binary population and then place it in a galactic
gravitational  potential to trace each binary until its
components merge due  to gravitational wave energy losses.
This method has to deal however with a number of uncertainties, that are
inherent in the binary population synthesis.
Moreover there are uncertainties
in what type and mass of a galaxy to use.

The binary population synthesis method has already been applied to the study
of compact  object binaries in the context of GRB progenitors. However
most studies have been concentrated only on double  neutron stars
and black hole neutron star systems.
Lipunov et~al. (1995) have used their ``scenario machine'' to model
the population of double neutron star  and black hole
neutron star binaries in a galaxy. They calculated the
expected $\log N$-$\log S$ GRB distribution assuming that they
are standard candles and compared it with
the BATSE observations.
Portegies-Zwart and Yungelson (1998) have considered
the origin and properties of double neutron star systems and black hole
neutron star binaries. They considered a few binary population synthesis
models, with varying kick velocities, initial binary
separations, initial mass ratios distribution and also considered cases 
with and without hyper accretion in the common envelope stage.
They found the rates of mergers to be consistent with the GRB rate provided that
GRBs are collimated to about ten degrees, and mentioned that double neutron stars
may travel Mpc distances out of a Milky Way like galaxy before merging.
Bloom, Sigurdsson \& Pols (1999) considered double neutron stars as
possible GRB progenitors, and calculated distributions of mergers of these binaries
around galaxies with different masses, varying the average 
kick velocities in the code. 
They found that a significant fraction of double neutron stars merge outside their 
host galaxies.
Bulik, Belczynski \& Zbijewski  (1999)
considered the mergers of binaries containing neutron stars, and
Belczynski, Bulik \& Zbijewski (2000) investigated differences between the
populations  of black hole neutron star binaries  and the
double neutron stars.     
Belczynski et al.\ (2000) found that black hole neutron star binaries
merge closer to the hosts than the double neutron stars.
Fryer et al.\ (1999a) considered other types of binary
progenitors of GRBs within the framework of the black hole accretion disk
model of the GRB central engine. These were white dwarf black hole mergers,
helium star black hole mergers, and collapsars in addition to
the double neutron star systems and black hole neutron star binaries.
They performed a thorough  parameter study, and repeated the calculations
with a number of modifications of their standard evolutionary model. They
calculated
the distribution of merger sites  in the potentials of galaxies
with the masses of $M_{MW}$ (Milky Way), $0.25 M_{MW}$, and  $0.01 M_{MW}$, 
however they do not vary the galactic size with mass.
Bloom, Kulkarni \& Djorgovski (2001)  presented a very detailed study of the
observational offsets between observed afterglows and 
GRB hosts galaxies. They compare these observations with the theoretical
distributions calculated with the code of Bloom et al.\ (1999)
and conclude that the so called delayed merging remnants i.e.
double neutron star systems and black hole
neutron star binaries are unlikely to be GRB progenitors, and argue in favor of
the prompt  bursters like collapsars and black hole helium star mergers.

In this work we extend our previous studies (Belczynski \& Bulik 1999;
Bulik et al.\ 1999; Belczynski et al.\ 2000) to include four more
proposed binary progenitors: compact object (black hole or
neutron star) white dwarf binaries and  Helium star mergers with black
holes or neutron stars.
We use a much improved and well tested binary  population
synthesis code and for consistency  we also present the updated
results for the two previously studied types of  proposed
progenitors: double neutron star and black hole neutron  star
systems. We calculate the properties of the ensemble of
each type of the proposed GRB progenitors and find their
distributions around different types of host galaxies.
We compare the observed GRB distribution around host galaxies
with the models.
In order to verify the robustness of the results we perform a
detailed parameter study and discuss the population synthesis
models which are responsible for the largest differences.

An additional  way of telling which group of the proposed binaries might be
responsible for GRBs is to predict their rates and compare
them to the observed rate of GRBs.  Population synthesis is a
powerful tool for predicting rates of  binary populations
although it suffers from many uncertainties as some parameters
of single and binary evolution are poorly known.  Moreover,
population synthesis works well in predicting  the relative
numbers of events, while calculation of absolute rates
requires additional assumptions.
However, such attempts have been made by a number of authors
mentioned above.  
Using the population synthesis method we calculate merger rates of 
white dwarf neutron star, white dwarf black hole, double neutron star, 
neutron star black hole systems, and formation rates of Helium star 
black hole and neutron star mergers. 
We compare the BATSE detection rate of GRBs with the cosmic rates of
the binary progenitors predicted in our calculations.

In \S\,2 we describe the population synthesis code {\em StarTrack} 
used to calculate
properties of binary GRB progenitors, in \S\,3 we present the
results, and finally \S\,4 is devoted to discussion with
conclusions.

\section{POPULATION SYNTHESIS MODEL}

\subsection{Stellar Evolution}

We use the {\em StarTrack} population synthesis code
(Belczynski, Kalogera \& Bulik 2001b).
Here, we summarize only the basic assumptions and ideas 
of the code.

The evolution of single stars is based on the analytic formulae 
derived by Hurley, Pols \& Tout (2000). 
With these formulae we are able to calculate the evolution of 
stars for Zero Age Main Sequence (ZAMS) masses: $0.5 - 100\,
{\rm M}_\odot$ and for metallicities:  $Z=0.0001 - 0.03$.
We follow the stellar evolution  from ZAMS through different
evolutionary phases depending on the initial (ZAMS) stellar mass: 
Main Sequence, Hertzsprung Gap, Red Giant Branch, 
Core Helium Burning, Asymptotic Giant Branch, 
and for stars with  their hydrogen-rich layers stripped off:
Helium Main Sequence, Helium Giant Branch.
We end the evolutionary calculations at the formation of a stellar
remnant:  a white dwarf, a neutron star or a black hole.
There are two modifications to the original Hurley et al.
(2000) formulae concerning the treatment of (i) final remnant masses, 
and (ii) Helium-star evolution (see Belczynski et al.\ 2001b; 
Belczynski \& Kalogera 2001).

The {\em StarTrack} code employs Monte Carlo techniques to model the 
evolution of single and binary stars.
In this work we use {\em StarTrack} to evolve large ensemble of
stars, and calculate statistical properties of the binary GRB
progenitors.

A binary system is described by four initial parameters: the mass $M_1$ of the
primary (the component which is initially more massive),
the mass ratio $q$ between the secondary and the primary, the semi-major axis of the
orbit $A$, and the orbital eccentricity $e$.
Each of these initial parameters is drawn from a distribution
and we assume that these distributions are independent. More specifically,
the mass of the primary is drawn from the Scalo initial mass function (Scalo 1986),
\begin{equation}
\Psi(M_1)\propto M_1^{-2.7}
\label{eq:Mf}
\end{equation}
and within the mass range $M_1=5-100M_\odot$.
The distribution of the mass ratios is taken to be
\begin{equation}
\Phi(q)=1,\;\;\;\;\; 0\leq q\leq 1\,,
\label{fq}
\end{equation}
following Bethe \& Brown (1998).
The initial binary separations are assumed as in Abt (1993)
\begin{equation}
\Gamma(A)\propto\frac{1}{A}\;,
\label{eq:fA}
\end{equation}
and finally, the initial distribution of the binary eccentricity is taken
following Duquennoy \& Mayor (1991)
\begin{equation}
\eta(e)=2e
\label{eq:fe}\;,\;\;\;\;\;0\leq e\leq 1\;.
\end{equation}

As we are interested only in the systems capable of producing binary GRB 
progenitors, containing at least one NS or BH, we evolve only massive
binaries, with primaries more massive than 5 M$_\odot$.
During the evolution of every system we take into account the effects
of wind mass-loss, asymmetric SN explosions, binary interactions
(conservative/non-conservative mass transfers, common envelope phases)
on the binary orbit and the binary components.
We also include the effects of accretion onto compact objects in
common  envelope (CE) phases
(Brown 1995; Bethe and Brown 1998; Belczynski et al.\ 2001b) and rejuvenation 
of binary components during mass transfer episodes.
Once a binary consists of two stellar remnants (NS, BH, WD), we calculate
its merger lifetime, the time until the components merge due to
gravitational radiation and associated orbital decay.

The {\em StarTrack} code may be used in several tens of modes, 
allowing for the change of main evolutionary parameters and 
initial distributions.
In the following, together with the given above initial distributions,
we define standard evolutionary model, with the set of parameters,
thought to represent  our best understanding of stellar single and
binary evolution.
 (1) {\em Kick velocities.} Compact objects receive natal kicks, when they
     form in supernova explosions.
Neutron star kicks are drawn from a weighted sum of two Maxwellian
distributions with $\sigma=175$\,km\,s$^{-1}$ (80\%) and
$\sigma=700$\,km\,s$^{-1}$ (20\%) (similar to the 
one of Cordes \& Chernoff 1998).
For black holes formed via partial fall back we use smaller kicks,
but drawn from the same distribution as for NS.
The kick scales with the amount of material ejected in SN explosion
or inversely with the amount of falling back material
(i.e., bigger the fall back, smaller the kick).
For BH formed in direct collapse of massive stars, we do not apply
any kicks, as  no supernova explosion accompanies the formation of
such objects.
 (2) {\em Maximum NS mass.} We adopt a conservative value of 
$M_{\rm max,NS}=3$\,M$_\odot$ (e.g., Kalogera \& Baym 1996).
 The mass of a compact object is estimated based on the mass and evolutionary
 status of its immediate progenitor, and not on any a priori assumptions.
 Once the mass of a compact object is calculated, its type (either NS or BH)
 is set by the value of $M_{\rm max,NS}$.
 Thus the choice of $M_{\rm max,NS}$ does not affect the overall population
 of compact objects.
 However, it affects the rates and various properties of binary GRB
 progenitors, as some of their groups contain either NS or BH (see \S\,3).
 (3) {\em Common envelope efficiency.} We assume $\alpha_{\rm
CE}\times\lambda = 1.0$, where $\alpha$ is the efficiency with which
orbital energy is used to unbind the stellar envelope (e.g., Webbink 1984), 
and $\lambda$ is the measure of the central concentration of the giant 
(e.g., Dewi \& Tauris 2000);
 (4) {\em Non--conservative mass transfer.} In cases of dynamically stable
mass transfer between non--degenerate stars we allow for mass and angular
momentum loss from the binary (see Podsiadlowski, Joss, \& Hsu 1992),
assuming that the fraction $f_{\rm a}$ of the mass lost from the donor is
accreted to the companion, and the rest ($1-f_{\rm a}$) is lost from the
system with specific angular momentum equal to $2\pi j A^2$/$P$.
We adopt $f_{\rm a}=0.5$ (e.g., Meurs \& van den Heuvel 1989) and $j=1$
(e.g., Podsiadlowski et al.\ 1992);
 (5) {\em Star formation history.}
We assume that star formation has been continuous in the disk of a given
galaxy.
To assess the properties of the current population of GRB progenitors,
we start the evolution of a single star or a binary system $t_{\rm birth}$\
ago, and follow it to the present time.
The birth time $t_{\rm birth}$ is drawn randomly from the range
0--10 Gyr, which corresponds to continuous star formation rate within
the disk of our Galaxy (Gilmore 2001).
(6) {\em Initial Binarity.}
We assume a binary fraction of $f_{\rm bi}=0.5$, which means that for
any 150 stars we evolve, we have 50 binary systems and 50 single stars.
(7) {\em Metallicity.}
We assume solar metallicity $Z=0.02$.
(8) {\em Stellar Winds.}
The single-star models we use (Hurley et al. 2000) include the effects
of mass loss due to stellar winds.
Mass loss rates are adopted from the literature for different
evolutionary phases.
For H-rich massive stars on MS (Nieuwenhuijzen \& de Jager (1990); using $Z$
dependence of Kudritzki et al.\ 1989), for RG branch stars
(Kudritzki \& Reimers 1978), on AGB (Vassiliadis \& Wood 1993) and
for Luminous Blue Variables (Hurley et al.\ 2000).
For He-rich stars W--R mass loss is included using the rates derived by
Hamann, Koesterke \& Wessolowski (1995) and modified by Hurley et al.
(2000).

\subsection{Dynamical Evolution of Stars in Model Galaxies}

The population synthesis code allows us to calculate the age for each 
system at the time when both stellar remnants have formed, and the 
subsequent merger time of a given system based on the remnant 
masses and their orbit.
We also calculate the systemic velocity gain due to asymmetric SN explosions 
and/or associated mass loss.
We use these informations to propagate binary GRB progenitor systems in
different galactic potentials, and to compute the distribution of their mergers
sites around different mass and size hosts. 

The potential of a spiral galaxy can be described as a sum of three
components: bulge, disk, and halo. 
A good representation of the galactic disk and bulge potential was
presented by Miyamoto \& Nagai (1975):
\begin{equation}
 \Phi(R,z) =
 {GM_i\over \sqrt{ R^2 + (a_i + \sqrt{ z^2 + b_i^2})^2}} 
\end{equation}
where the index $i$ refers to either bulge or disk, $a_i$ and $b_i$ 
are the parameters, $M$ is the mass, $R=\sqrt{x^2 +y^2}$ and the $x-y$
coordinates  span the galactic plane.  
The dark matter  halo potential is spherically symmetric
\begin{equation}
 \Phi(r) = - {G M_h\over  r_c} \left[ {1\over 2} 
 \ln\left( 1 + {r^2\over r_c^2}\right) + {r_c\over r} 
 {\mathrm{atan}} \left( r\over r_c\right) \right] 
\end{equation}
where $r_c$ is the core radius. 
The halo potential corresponds to a mass distribution 
$\rho = \rho_c/[1 + (r/r_c)^2]$, 
and we introduce a cutoff radius $r_{cut}$ beyond which 
the halo density falls to zero, in order to make the halo mass
finite and the halo gravitational potential is 
$\Phi(r)\propto r^{-1}$ when $r>r_{cut}$. 

We consider galaxies with  four masses, expressed in the units of Milky 
Way mass ($M_{\rm MW}=1.5 \times 10^{11} M_\odot$):
$1.0, \ 0.1,\ 0.01,\ 0.001 \times M_{\rm MW}$.
For the Milky Way mass galaxy the bulge potential ($i=1$) is described by: 
$a_1 =0\,$kpc, $b_1 = 0.277\,$kpc, $M_1 = 1.12\times 10
^{10}\,M_\odot$; 
the disk potential ($i=2$): 
$a_2 = 4.2\,$kpc, $b_2 = 0.198\,$kpc,  $M_2 = 8.78\times 10^{10}\,M_\odot$; the
halo potential: $r_c =6.0\,$kpc, and $M_h = 5.0 \times 10^{10}\,M_\odot$ and
the $r_{cut}=100$kpc (Paczynski 1990; Blaes \& Rajagopal 1991). 
To obtain the  potential of a galaxy with the mass $\alpha M_{MW}$ we rescale 
all the masses by the factor of $\alpha$ and the distances $a_i, b_i, r_c, R, z$
by $\alpha^{1/3}$.
Such scaling keeps the galaxy density constant, and we made sure that our model 
galaxies have flat rotation curves.

We adopt the following distribution
of stars 
within the disk of a given galaxy (Paczynski 1990):
$ P(R,z)dRdz=P(R)dRp(z)dz.$
The radial distribution is exponential
\begin{equation}
P(R) dR \propto R \,e^{-R/R_{exp}} dR
\end{equation}
and extends up to $R_{\rm max}$.
 The vertical distribution is also exponential
\begin{equation}
p(z) dz \propto e^{-z/z_{exp}} dz.
\end{equation}
For a Milky Way type galaxy we have $R_{exp}=4.5 {\rm kpc}, R_{\rm max}=20
{\rm kpc}, z_{exp}=75 {\rm pc}$, and these parameters are assumed to scale 
with the galaxy mass as $\alpha^{1/3}$.

Each binary moves initially with the local rotational velocity
in its galaxy, and has no vertical component of velocity.
After each supernova explosion the kick imparted on the binary is added
and the binary trajectory is calculated until the merger occurs.

\section{RESULTS}

\subsection{Binary GRB Progenitor Types}

Fryer et al. (1999b), suggested the possibility that white dwarf neutron
star (WD-NS) merger may lead to formation of a black hole accretion disk 
system, followed by a GRB.
Since the GRB outburst mechanism is not well understood, the 
results of hydrodynamic calculations of stellar mergers should be 
treated with some caution.
However, we will consider the group of WD-NS systems as  potential GRB 
progenitors, for the sake of  completeness of the study. 
From the entire group of coalescing WD-NS binaries we chose these,
 which have
the best chance to produce observable GRBs, i.e., systems in which WD are
more massive than $M_{min,WD}=0.9 M_\odot$ to make sure that the
mass transfer onto the NS is unstable, and the total mass of
the system satisfies
$M_{WD}+M_{NS}> M_{max,NS}+0.3\,M_\odot$, since
we require that the NS has to accrete enough material to collapse
to a BH and
that the disk formed in the merger must have the mass of at least
$\sim 0.3\, M_\odot$ to produce a GRB (Fryer et al. 1999b, 2001).

Fryer et al. (1999b) also suggested that mergers of white dwarf black hole
(WD-BH) binaries may give a rise to GRBs.
Following the Fryer et al. (1999b) hydrodynamical calculations, we require that
WD mass is larger than $0.9 M_\odot$ to classify a coalescing WD-BH system
as a potential GRB progenitor.
Only for these high WD masses, the accretion of WD onto BH is dynamically
unstable and the rapidly disrupted (in several binary rotations) WD forms
a thick
disk around BH, which may give a rise to GRB.

\begin{figure*}[t]
\centerline{\psfig{file=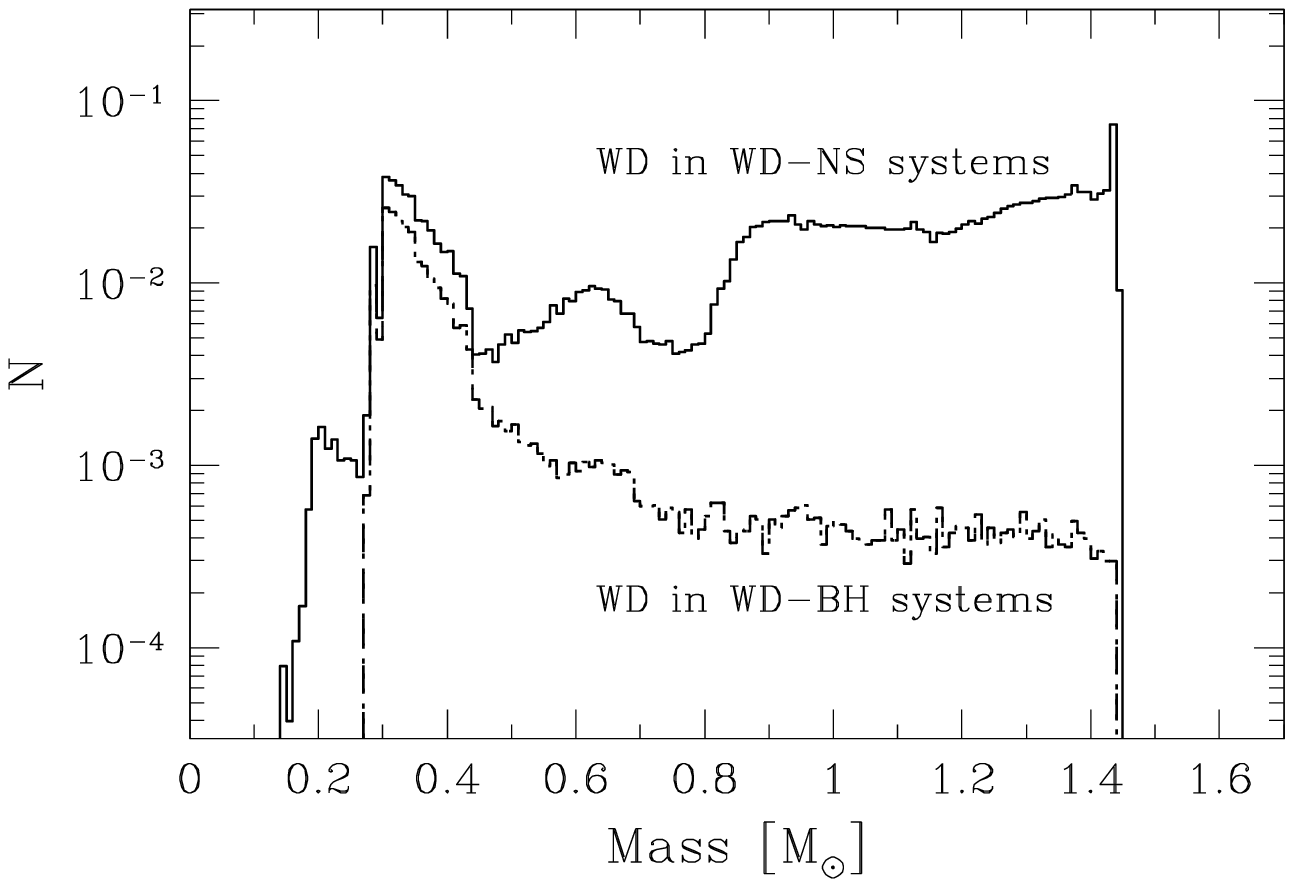,width=0.9\textwidth}}
\caption{White dwarf mass distributions in WD-NS (solid line) and
WD-BH systems (dashed line) for our standard evolutionary model.
Distributions are normalized to the total number of binary GRB
progenitors (100800) formed out of $N_{\rm TOT}=3 \times 10^7$
primordial binaries.
We require that a WD mass exceeds $0.9 M_\odot$ for a given
system to be classified as a potential GRB progenitor.
We also study models in which the minimum mass is 0.7 and 1.1 
$M_\odot$. 
}
\label{wddis}
\end{figure*}

In Figure~\ref{wddis} we present the
distribution of WD masses in coalescing WD-NS and
WD-BH systems.
For WD-BH binaries, WD masses distribution rises sharply at $\sim 0.3
M_\odot$ and then falls approximately exponentially to flatten out
for masses higher that $\sim 0.7 M_\odot$.
Therefore, changing the $M_{min,WD}$ to slightly higher/smaller values,
will decrease/increase the number of WD-BH GRB progenitors roughly
proportionally to the WD mass limit change.
For WD-NS systems, the WD masses distribution is also rather flat close to
and over $M_{min,WD}=0.9 M_\odot$.
However, as the value of limiting mass of WD is highly uncertain, we will
present two models with the decreased and increased $M_{min,WD}$.

Fryer \& Woosley (1998) proposed yet another type of  binary GRB progenitors,
i.e. binaries merging in CE events, with one component 
being an evolved (giant) star 
and the other already a compact object, either NS or BH.
In this scenario, the binary does not have enough
orbital energy to eject the common envelope, so the compact object spiraling in
finally merges with the helium core of the giant.
The compact object disrupts tidally the helium star, accreting part of its
material, and becoming a BH, if it was not one already.
The remainder of the 
giant's helium core forms a thick accretion disk around the BH,
a configuration, which is believed to give a rise to a GRB.

In our models we distinguish systems  that
contain either a NS or a BH at the 
onset of the CE phase, leading to the final merger.
We will denote systems containing NS as helium star/neutron star mergers 
(He-NS), and containing BH as helium star/black hole mergers (He-BH). 
Following the detailed studies of He-NS and He-BH mergers (Bottcher \& Fryer
2000; Zhang \& Fryer 2000), we choose  only these 
systems in which helium core mass exceeds $M_{min,He}=6 M_\odot$
as GRB progenitors.

\begin{figure*}[t]
\centerline{\psfig{file=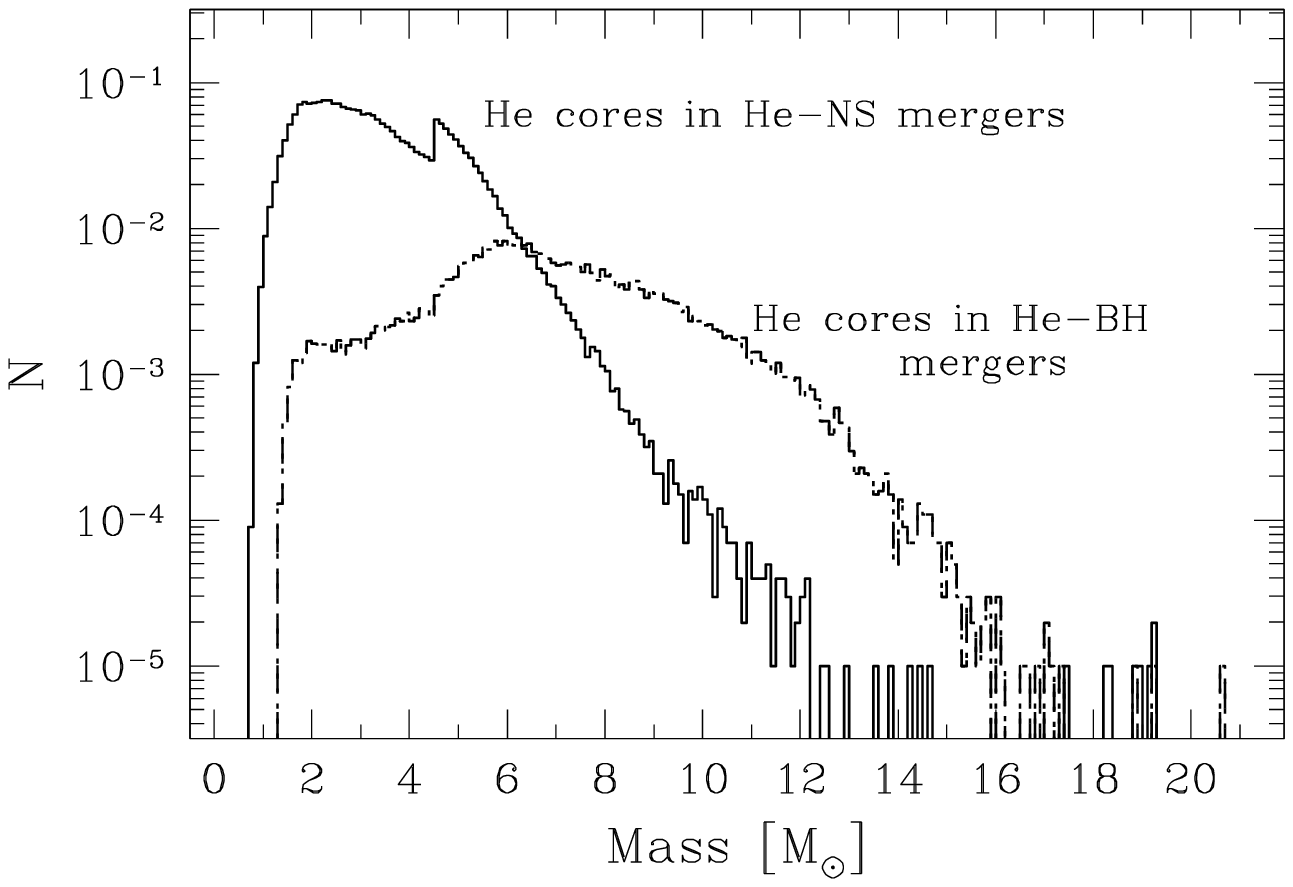,width=0.9\textwidth}}
\caption{Helium core mass distributions in He-NS (solid line) and 
He-BH mergers (dashed line) for our standard evolutionary model.
Distributions are normalized to the total number of binary GRB 
progenitors (100800) formed out of $N_{\rm TOT}=3 \times 10^7$ 
primordial binaries.
We require that a He core mass exceeds $6 M_\odot$ for a given 
system to be classified as a potential GRB progenitor.
We also study models in which the minimum mass is 4 and 8 
$M_\odot$.
}
\label{hedis} 
\end{figure*}

The distributions of helium core masses for He-NS and He-BH at the onset of 
the final CE phase are shown in Figure~\ref{hedis}.
Both distributions rise sharply at $\sim 1 M_\odot$, then for He-NS mergers
the distribution falls down rapidly above  $5 M_\odot$, while 
 for He-BH mergers 
the distribution  decrease  starts at  a higher mass, 
$7 M_\odot$, and is more gradual.
Thus the number of GRB progenitor He-NS mergers depends strongly on
the  value of $M_{min,He}$, so we will present models with
different value of this limiting helium core mass.  

Finally, the most intensively studied binary progenitors of GRBs
are double neutron 
stars (NS-NS) and neutron star black hole systems (NS-BH).
In defining the boundary between NS-NS systems and NS-BH systems
we assume that the maximal mass of a neutron star is $M_{\rm max,NS}=3 \,
M_\odot$, however we will also present results for two smaller limiting 
masses of $2.0, 1.5 \, M_\odot$.
The mass distribution of compact objects in these types of binaries, starts 
in our code with the maximum at the $\sim 1.2 M_\odot$, 
followed by a rapid decline (which
reflects the shape of the assumed initial mass function) and then at around 
$3 M_\odot$ flattens out and stays roughly constant up to the 
highest BH masses of
$\sim 14 M_\odot$.
The maximum BH mass is set by the effect of wind mass loss on massive stars.
This distribution is presented and discussed in detail by Belczynski 
et al.\ (2001b).

Belczynski \& Kalogera (2001) and Belczynski, Bulik 
\& Kalogera (2001a) identified new subpopulations of NS-NS binaries.
The new subpopulations dominate the group of coalescing NS-NS systems, and
moreover, they were found to exhibit quite different properties than 
the systems studied to date.
Given the importance of these subpopulations to our conclusions,
in the following subsection we briefly summarize the results of
Belczynski \& Kalogera (2001) and Belczynski et al.\ (2001a).

\subsection{Double neutron star binaries}

Double neutron stars are formed in various ways, including more than
14 different evolutionary channels, identified in Belczynski et al.\
(2001b).
We find that the entire population of coalescing NS-NS systems, may be divided
into three subgroups.

{\em Group I} consists of non-recycled NS-NS systems (containing two
non-recycled pulsars), which finish their
evolution in a double CE of two helium giants.
Two bare CO cores emerge after envelope ejection, and they form
two NS in two consecutive SN type Ic explosions.
Provided that the system is not disrupted by SN kicks and mass loss,
the two NS form a tight binary, with the unique characteristic
that none of the NS
had a chance to be recycled.
For more details see Belczynski \& Kalogera (2001).
{\em Group II} includes all the systems that finished their evolution
through a CE phase, with a helium giant donor and a NS companion.
During the CE phase a NS accretes material from the  envelope of
the giant, becoming most probably a recycled pulsar.
The carbon-oxygen core of the Helium giant forms another NS soon after 
the CE phase ends.
The system has a good chance to survive even if  the newly born NS
receives a high kick because it is very tightly  bound after the 
CE episode.
For more details see Belczynski et al. (2001a).
{\em Group III} consists of all the other NS-NS systems formed,
through the classical channels (e.g. Bhattacharya \& van den Heuvel 1991).

In our standard model, group II strongly dominates the population 
of coalescing NS-NS systems (81\%) over group III (11\%) and I (8\%). 
This is due to the fact that we allow for helium star radial evolution, and
usually just prior to the formation of a tight (coalescing) NS-NS system we
encounter one extra CE episode, as compared to the classical channels.
This has major consequences for the merger time distribution of the NS-NS
population, and in turn for the distribution of NS-NS merger sites around
their
host galaxies.    
Merger times of classical systems are comparable with the 
Hubble time, and that
gives them ample time to escape from their host galaxies.
As it has been shown in previous studies (e.g., Bulik et al.\ 1999; 
Bloom et al.\ 1999) which did not include Helium
star detailed radial evolution, a significant fraction of the NS-NS
population tended to merge outside host galaxies, exactly like
group III - the 
classical systems.
In contrast, the binaries of Groups I and II, due to the extra CE episode, are
tighter, and their merger times are much shorter: of order of $\sim 1$\,Myr.
Thus even if they acquire high systemic velocities due to the asymmetric SN
explosions, they will merge within the host galaxies, near the places they
were born.        
Group I and II dominate the population, and thus the overall NS-NS
distribution of merger sites will follow the distribution of primordial
binaries or star formation regions in the host galaxies.

The formation of NS-NS systems of group I and II depends on the
assumption that evolved low mass helium donors can initiate and survive 
CE phase.
This assumption have yet to be proven by detailed hydrodynamical
calculations. 

Since, the properties of these new subpopulations were already 
discussed separately (Belczynski et~al. 2001a) we present here 
only the results for the overall population of NS-NS binaries.

\subsection{Characteristic Binary Timescales}

\begin{figure*}[t]
\centerline{\psfig{file=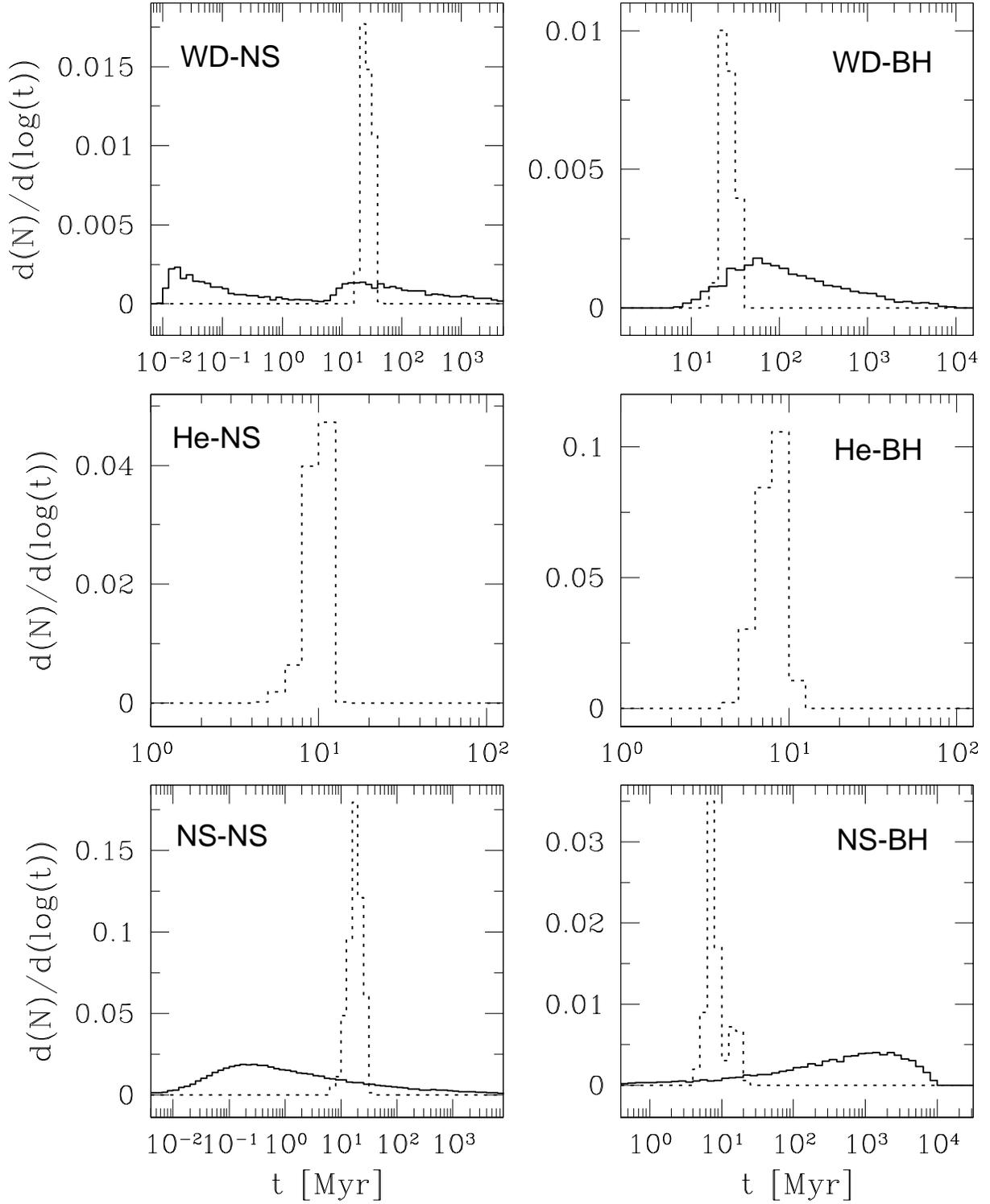,width=0.9\textwidth}}
\caption{Distributions of evolutionary (broken lines) and merger 
(solid lines) times for GRB binary candidates calculated in our 
standard evolutionary scenario.
Distributions are normalized to the total number of binary GRB
progenitors (100800) formed out of $N_{\rm TOT}=3 \times 10^7$
primordial binaries.
Note that every panel has diffrent vertical and horizontal scales.
}
\label{times}
\end{figure*}

We call the time a given system needs to evolve from ZAMS 
to form two stellar 
remnants, the evolutionary time and denote it by $t_{\rm evol}$. 
We call the time  required for these stellar remnants to merge due to
gravitational radiation the merger time and denote it by $t_{\rm merg}$.
The total lifetime of a given system is the sum of the two:
$
t_{\rm life} = t_{\rm evol} + t_{\rm merg}\,$.
In Figure~\ref{times} we show the distributions of both evolutionary and
merger times for all GRB binary candidates, while in Table~\ref{Times} 
we list the distributions medians and spans defined as 
the  time range containing
 90\% of the systems around the median.

In general, the evolutionary delays are of the order of a few to several
tens Myrs, and their distributions are rather narrow for different types 
of systems.
Since the rate of evolution of a given star depends primarily on its mass,
$t_{\rm evol}$ is set mainly by the mass of a given binary components.
The evolution proceeds  slower for less massive stars,
and $t_{\rm evol}$ is determined in general by the mass of the 
secondary (unless the mass ratio is reversed due to the mass transfer).
This is why $t_{\rm evol}$ for WD-NS and WD-BH systems is the longest
and almost equal ($\sim 26$\ Myr), as WD are the lightest components of
binary GRB candidates.
The  NS-NS and NS-BH binaries are formed in shorter times,
with the median
of $t_{\rm evol}$ distributions of $\sim 19$\ and $\sim 8$\ Myr,
respectively.
Evolutionary times for He-NS and He-BH $t_{\rm evol}$ are very short
($\sim 9$\ Myr), as they finish their evolution even before formation of
a second remnant.

Merger times are quite different for various binary GRB candidates.
For He-NS and He-BH mergers we do not list $t_{\rm merg}$, as these events
take place even before two stellar remnants are formed, due to the components 
merger in CE spiral in.
The shortest merger times are found for NS-NS binaries and for WD-NS
systems, with the medians of $\sim 0.7$\ and $\sim 6.8$\ Myr, respectively.
Much longer merger times are characteristic of WD-BH systems ($\sim 97$\
Myr), with the longest $t_{\rm merg}$ found for NS-BH binaries
($\sim 535$\ Myr).

\subsection{Event Rates}

The method of  population synthesis requires use of quite a number of
parameters, and initial distributions of variables which may
affect the final results. In order to assess their influence
on the final results we have repeated the calculations with varying
evolutionary parameters. 
The models and the differences with the standard
model are listed in Table~\ref{models}.
The coalescence rates of different types of GRB progenitors
within each model
are shown in
Table~\ref{rates}.
They have been calibrated to the Type II supernova empirical
rates and normalized to our Galaxy (Capellaro, Evans \& Turatto 1999).
The standard model (A) results are based on a simulation
of $3\times 10^7$ binaries, while the remaining models are the simulations
of at least $10^6$ binaries.  The statistical accuracy of most rates is
better than a few percent, however in some cases where the rates are smaller
than $1\,$Myr$^{-1}$ the accuracy is of the order of a few tens percent,
yet improving them would require a huge computational effort.

Models B1-13 represent the results of evolution with different kick
velocities imparted on the  compact objects.
In model B1 we assume symmetric SN explosions, whereas in models B2--12
we draw the kick velocity $V_k$ from a single Maxwellian:
\begin{equation}
g(V_k) \propto {V_k}^2 \exp\left[{-({V_k} / \sigma)^2}\right],
\end{equation}
varying $\sigma$ values in  the range $10-600\ {\rm km s}^{-1}$.
In model B13 we use a kick distribution of the form suggested by
Paczynski (1990):
\begin{equation}
f(V_k) \propto [1+(V_k/\sigma)^2]^{-1},
\end{equation}
which allows
for a significant fraction of low--magnitude kicks. We use $\sigma = 600\ 
{\rm km s}^{-1}$, which gives a reasonable fit to the population of single
pulsars in the solar vicinity (Hartman 1997).

\begin{figure*}[t]
\centerline{\psfig{file=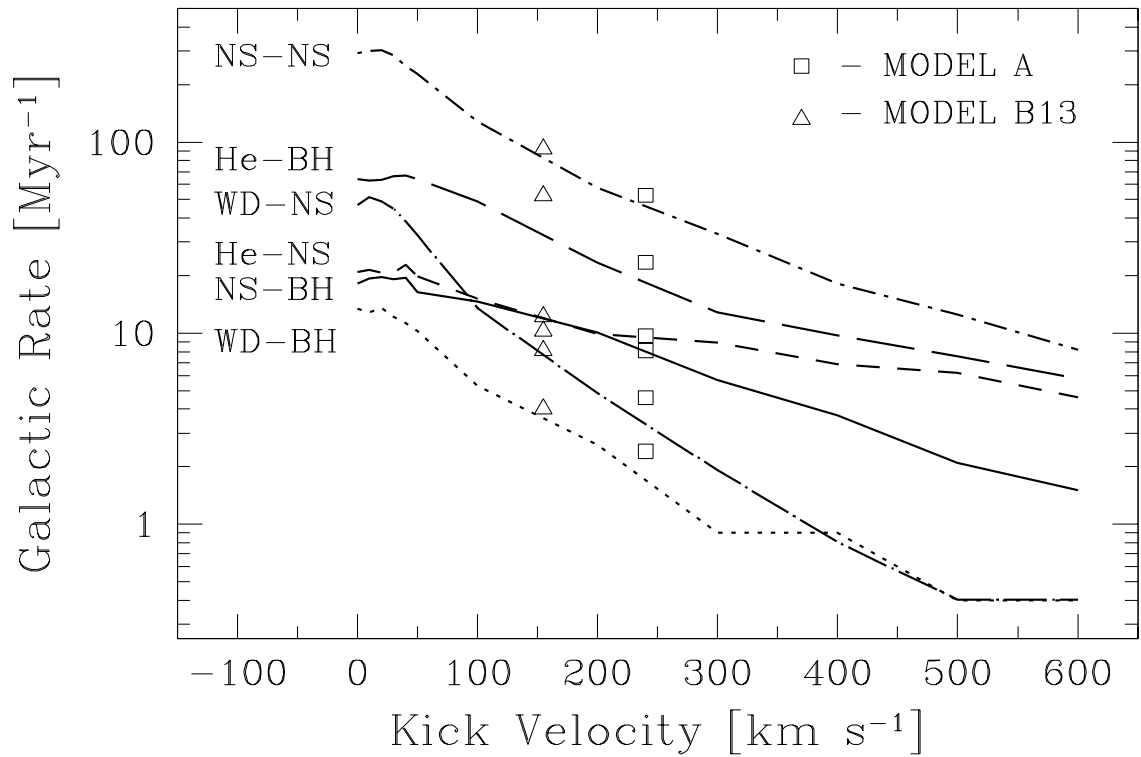,width=0.9\textwidth}}
\caption{The dependence of galactic GRB progenitor coalescence rates on the 
assumed natal kick velocity distribution.
Lines connect rates for models B1-B12 and the horizontal scale shows the
width of Maxwellian kick distribution of a given model.
Triangles mark rates of our standard model (A) and squares
he model with `Paczynski''
kick distribution (B13).
The width of kick velocity distribution scale is irrelevant for these
two models, and they were placed in horizontal axis to approximately
match the rates obtained with single Maxwellian kick velocity 
distribution.
}
\label{frac}
\end{figure*}

In Figure~\ref{frac} we show the dependence of
coalescence rates on the assumed kick velocity distribution.
Due to the  fact that high kicks tend to disrupt binaries,
the number of  systems formed with at least one compact object,
falls off quickly with the kick velocity.
This has already been noticed for NS-NS and NS-BH binaries (e.g., Lipunov,
Postnov \& Prokhorov 1997; Belczynski \& Bulik 1999), and also for 
WD-BH systems and He-BH mergers (Fryer et al. 1999a).
As seen from Figure~\ref{frac}, the coalescence rates of 
GRB progenitors fall off 
approximately exponentially with the kick velocity.
Note however, that the slope is smaller for He-NS and He-BH mergers
than for other types of progenitors. 
This is due to (i) only one kick the system receives, (ii) the relatively
high total mass of the binary (recall that $M_{min,He}=6.0 M_\odot$), so 
the kick imparted to NS or BH does not have a big impact on such
systems.
    
For the majority of models the rates for WD-NS mergers
stay close to several coalescence events per Myr per Galaxy.
However, the rate changes significantly ($0.03-114.7$\ Myr$^{-1}$) for 
a few extreme models.
Besides the
 strong dependence  of the rates on kick velocity  discussed above,
a large number of WD-NS is produced in models D1 and D2, in which
the neutron star maximal mass is smaller than in the other models.
This increase is due to our requirement that only systems with total
mass higher than the maximal neutron star mass increased by 0.3 $M_\odot$
are  classified as GRB progenitors.

In fact, the coalescence rate of {\em all} WD-NS systems 
(irrespective of WD or total system mass) is as high as 
$204.5$\ Myr$^{-1}$ and $128.9$\ Myr$^{-1}$ for our standard 
model and model D2, respectively.
For the very low CE efficiency of model E1, the rate drops down almost to
zero, since
in this model many systems potentially able to
form a WD-NS GRB progenitor evolve
through CE phase.
Once the CE efficiency drops, a system needs to use more orbital energy to
expel the envelope, and it becomes tighter.
At very small efficiencies, there is not enough orbital energy for envelope
ejection, and the two stars merge, thus decreasing the final number of
WD-NS systems.

Coalescence rates of WD-BH vary much less than these of WD-NS
progenitors, and they stay close to few coalescence events per Myr
per Galaxy for most models.
The largest change ($0.2-22.1$\ Myr$^{-1}$) appears  for  models D2 and E1
similarly to the case of WD-NS.
The lowest rate of model E1 is explained in the same way as for WD-NS
progenitors.
The highest rate of model D2, reflects the fact of the lowest NS/BH mass
limit, and  many systems classified in other models as WD-NS, here are
counted as WD-BH.

The coalescence rates of He-NS change almost by 3 orders of magnitude
($0.1-73.6$\ Myr$^{-1}$), although for most models, including the standard one, they
remain close to several events per Myr per Galaxy.
The number of helium star mergers strongly (2 orders of magnitude)
depends on the required  minimum mass of the Helium core - see
models P1-2 and also Figure~\ref{hedis}.
This is explained by our adopted IMF, which gives more low mass stars,
therefore low-mass helium cores are much more abundant
(note the high rate of model P1 with $M_{min,He}=4.0 M_\odot$)
than the massive ones (low rate of model P2 with $M_{min,He}=8.0 M_\odot$).
The lowest rates are found for models D1 and D2, with the maximal
NS mass of 2.0 and 1.5 $M_\odot$.
With decreased $M_{max,NS}$, as compared to our standard model, we choose
only the lightest possible primaries, which will evolve to form NS.
On the other hand we require that the secondary must form $6.0 M_\odot$
helium core, so it needs to have been massive already at the start.
In these models only the binaries with relatively comparable mass components
($q \sim 1$) may evolve to form He-NS mergers.
Thus, decreasing $M_{max,NS}$, narrows down the range of $q$ in which He-NS
mergers may be formed, which results in drop of their rate (recall that
we adopted a flat initial mass ratio distribution).

The rates for He-BH mergers are the most independent of model parameters
varying just by an order of magnitude ($4.5-91.6$\ Myr$^{-1}$), which considering
the extreme changes in model parameters and initial
distributions is quite remarkable.
The coalescence rate for standard model is 23.5 events per Myr per Galaxy, 
and it remains approximately at this level for the majority of models.

The dependence of NS-NS and NS-BH merger rates on the model parameters
is discussed in detail by Belczynski et al.\ (2001b), and in what follows
we restrict the description just to a brief summary.

Merger rates for NS-NS systems change by two orders of magnitude
in various models
($2.5-302.2$\ Myr$^{-1}$) and  the  standard model rate is about 50 merger events
per Myr per Galaxy.
As these systems experience two SN explosions, and the NS receive highest
possible kicks (not lowered as in the case of BH), their rate depends
very strongly on the assumed kick velocity distribution.
The highest rates are found for smallest kick models (B1-5).
Production of coalescing NS-NS binaries is greatly reduced by
reducing CE efficiency (model E1), for the reasons described above
(see the discussion of WD-NS merger rates).
Also altering the distribution of the initial mass ratio (model M2)
changes the rates significantly and leads to an enhanced production of NS-NS
systems.

Finally, the merger rate of NS-BH systems stays at a rather constant 
level ($1.3-36.2$\ Myr$^{-1}$),
with most model rates of approximately 10 events per Myr per Galaxy.
The rate is not so sensitive to the kick velocity as the merger rates of
NS-NS systems, because NS-BH binaries receive at least one smaller kick
(that imparted on the BH) and also NS-BH systems are more massive,
so the kicks have smaller chance to disrupt them.
The smallest merger rate is found for the model with the enhanced wind
mass loss rate (model G2).
Due to the high mass loss the stars do not form massive compact
objects, and the number of BH formed (and systems harboring BH) is
greatly reduced.
The highest rate is achieved by the shift in NS maximal mass of model
D2 to its lowest value adopted here, which enhances the
 rate of NS-BH and
depletes the rate of NS-NS systems.

\subsection{Redshift distribution}

\begin{figure*}[t]
\centerline{\psfig{file=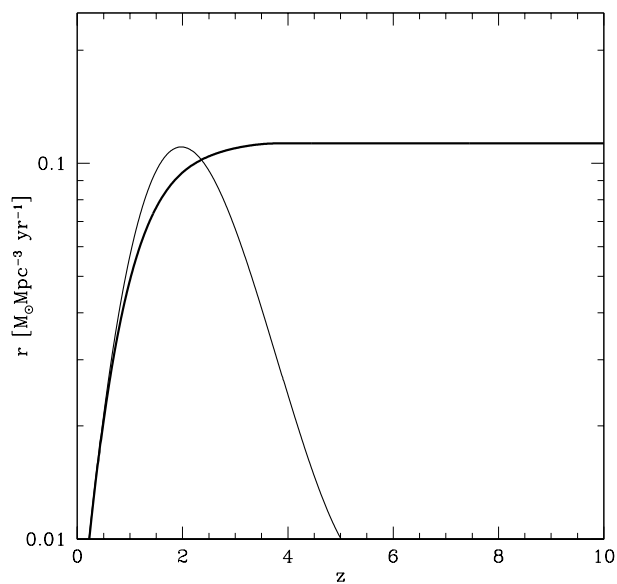,width=0.45\textwidth}}
\caption{
Star formation history rates used in this work. The thin
line is based on lower limits from Madau et al.\ (1996), 
while the thick line represents approximately the rate of 
Rowan-Robinson (1999).
}
\label{sfr}
\end{figure*}

{\em Standard Model}
The results of the population synthesis code can be combined
with  the cosmic star formation rate history to
yield the  rate of various types of GRB progenitors as a function
of redshift.
Star formation history at high redshift is not well
known, however it is generally agreed that the star formation
rate rises steeply up to  $z\approx 1$. At higher redshifts  the
analysis of the Hubble Deep Field  (Madau et al.\ 1996) provided
lower limits on the rate, yet these limits decrease with
increasing redshift. On the other hand Rowan-Robinson (1999) argues
that the star formation does not decrease and remains roughly at
the same level above $z=1$.
We consider two cases: a star formation function falling
down steeply above $z\approx 1$ (the thin line in
Figure~\ref{sfr}), and a case of strong star formation continuing
up to $z=10$ (the thick line in Figure~\ref{sfr}).
We adopt a flat cosmology model with density parameter of matter
$\Omega_m=0.3$, density parameter of cosmological constant 
$\Omega_\Lambda=0.7$ and for Hubble constant $H_0=65 {\rm km} 
\ {\rm s}^{-1} {\rm Mpc}^{-1}$.

For a given type $i$ of the GRB progenitor we can calculate
the number of events up to the redshift $z$ per unit of observed time:
\begin{equation}
rate_i(<z) = 4\pi \int_0^z r_z^2 {dr_z\over dz} {R_i(z)\over 1+z}
dz\, ,
\end{equation}
where $r_z$ is the effective distance $r_z = c H_0^{-1}\int_0^z
(\Omega_m(1+z^3)+\Omega_\Lambda)^{-1/2} dz$, and $c$ is the speed of light.
$R_i(z)$ is the rate of a given type of event at the redshift of $z$:
\begin{equation}
R_i(z) = \int_{t(z)}^{t(z=\infty)}  R_{sfr}(t') \,f_i\, p_i(t(z)-t') dt' \, ,
\end{equation}
where $t$ is the dynamical time, $dt = -H_0^{-1} (1+z)^{-1}
(\Omega_m(1+z^3)+\Omega_\Lambda)^{-1/2}dz$,
$p_i(t)$ is the probability density of a merger of a given type
as a function of time since formation of the system, and
$f_i$ is the mass fraction of the  binaries in the
entire stellar population (single and binary) of mass range
[0.08--100 ${\rm M}_\odot$] that can lead to formation of GRB
progenitors of type $i$.
$R_{sfr}(t)$ is the cosmic star-formation rate at a time $t$ or a
corresponding redshift $z$.
We obtain the probability density $p_i(t)$ numerically for each type of a merger
using the  population synthesis code.
In calculation of $p_i(t)$ we take into account both, the evolutionary time
delay (from formation of the system until two components form stellar
remnants) and the merger time delay (the time needed for two stellar
remnants to merge due to gravitational wave emission).

\begin{figure*}[t]
\centerline{\psfig{file=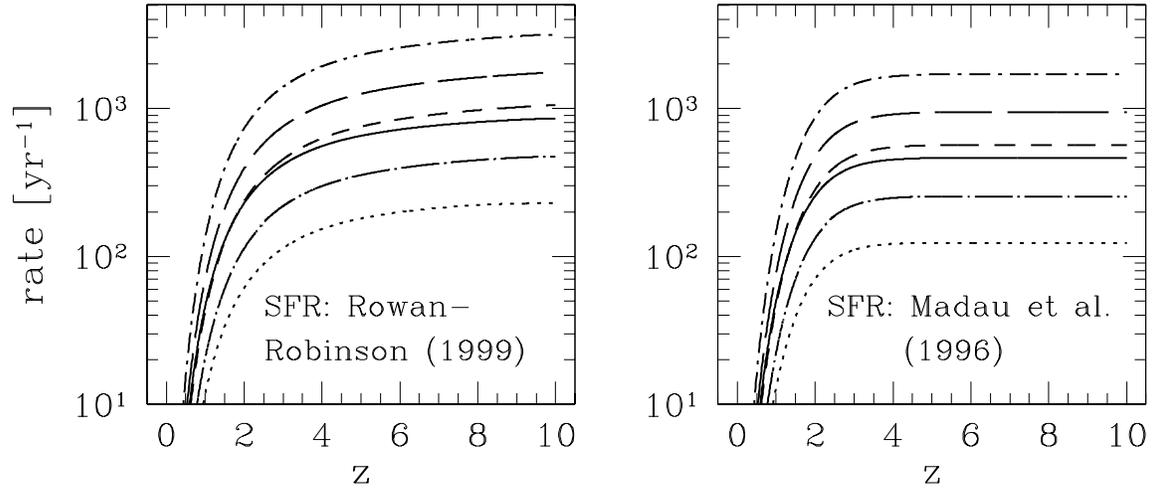,width=0.9\textwidth}}
\caption{
Cumulative event rates of different GRB progenitor types as a function of
redshift for our standard evolutionary model.
From top to bottom curves correspond to: NS-NS (dotted--short dashed line), 
He-BH (long dashed line), He-NS (short dashed line), 
NS-BH (solid line), WD-NS (doted--long dashed line) and WD-BH 
mergers (dotted line).
The left panel shows the case with   assumed star formation rate history of
Rowan-Robinson (1999), while the right panel to that of Madau et al.\ 
(1996).
For all calculations flat cosmology model was used, with 
$\Omega_{\rm m}=0.3$ and $\Omega_{\Lambda}=0.7$. 
}\label{cosmo}
\end{figure*}

The redshift dependence of GRB progenitor rate is presented for our
standard model and for the two adopted star formation rate histories 
in Figure~\ref{cosmo}.
For any given $z$ the GRB progenitor merger rates are the highest 
for NS-NS binaries, than for He-BH and He-NS mergers, which are 
closely followed by the NS-BH systems.
We find the lowest rates for mergers of WD-NS and WD-BH binaries.

The shape of the 
star formation rate determines the shape of the GRB progenitor rate 
redshift distribution.
For the Rowan-Robinson (1999) SFR,  
progenitors are expected even at very high
redshifts ($z \sim 10$), while for the SFR of Madau et al.\ (1996) we do not 
expect to produce any GRBs from binary mergers over $z \la 4$.
However, GRBs are observed at high redshifts.
The highest spectroscopic redshift $z=4.500 \pm 0.015$ was measured for 
GRB 000131 (Andersen et al.\ 2000), while 
Fruchter et al.\ (1999) estimated photometrically the redshift of GRB 980329 
to be $\simeq 5$ (although following Bloom et al.\ 2001, in the
Table~\ref{Loc} we list for this burst a more moderate estimate of $z \la 3.5$).
Therefore, if we assume that GRBs originate in binary progenitors our results
argue against the SFR drawn along the lower limits of
Madau et al.\ (1996), while GRBs observed at high
redshifts are in agreement with our results based on SFR of
Rowan-Robinson (1999).

The curves in Figure~\ref{cosmo} can  be  compared with the
BATSE gamma-ray burst detection rate  corrected for BATSE sky
exposure, which is $\approx 800$ events per year. 
Only the rate of NS-NS and He-BH mergers is significantly above the
BATSE observed rate, if we count the merging events up to the highest
GRBs observed redshifts of $z=4-5$.
The He-NS and NS-BH merging rates in our standard
model are marginally consistent
with the observed rate, and these only for the  Rowan-Robinson (1999) SRF model.
Within the standard population synthesis model 
progenitors with WD merge at considerably lower rates than that
expected for GRB progenitors.
The predicted cumulative rates presented in Figure~\ref{cosmo} 
will  decrease if we account for collimation and thus restricted 
visibility of gamma-ray bursts.
Since GRBs are thought to be collimated (Harrison et al.\ 1999;
Stanek et al.\ 1999; Kuulkers et al.\ 2000; Panaitescu \& Kumar 
2001) this puts further limits on the binary progenitors.
If any degree of collimation is taken into account,
we may also argue against NS-BH and He-NS mergers besides
WD-NS and WD-BH binaries
as the sole progenitors of GRBs.
Moreover, for most frequent NS-NS and He-BH mergers it would be difficult
 to reproduce the observed GRB rate with any significant  degree of
GRB collimation.
The total rate of {\em all} binary mergers is $\sim 7000$ per year 
(up to $z=5$), and
the collimation which would reduce this number to the observed BATSE
rate would be $\Theta \approx 25^\circ$ (the outflow half-opening angle).
Of course, if any single binary merger model was to reproduce the
observed rate, the required collimation would be much smaller (i.e.,
$\Theta$ much bigger).

\begin{figure*}[t]
\centerline{\psfig{file=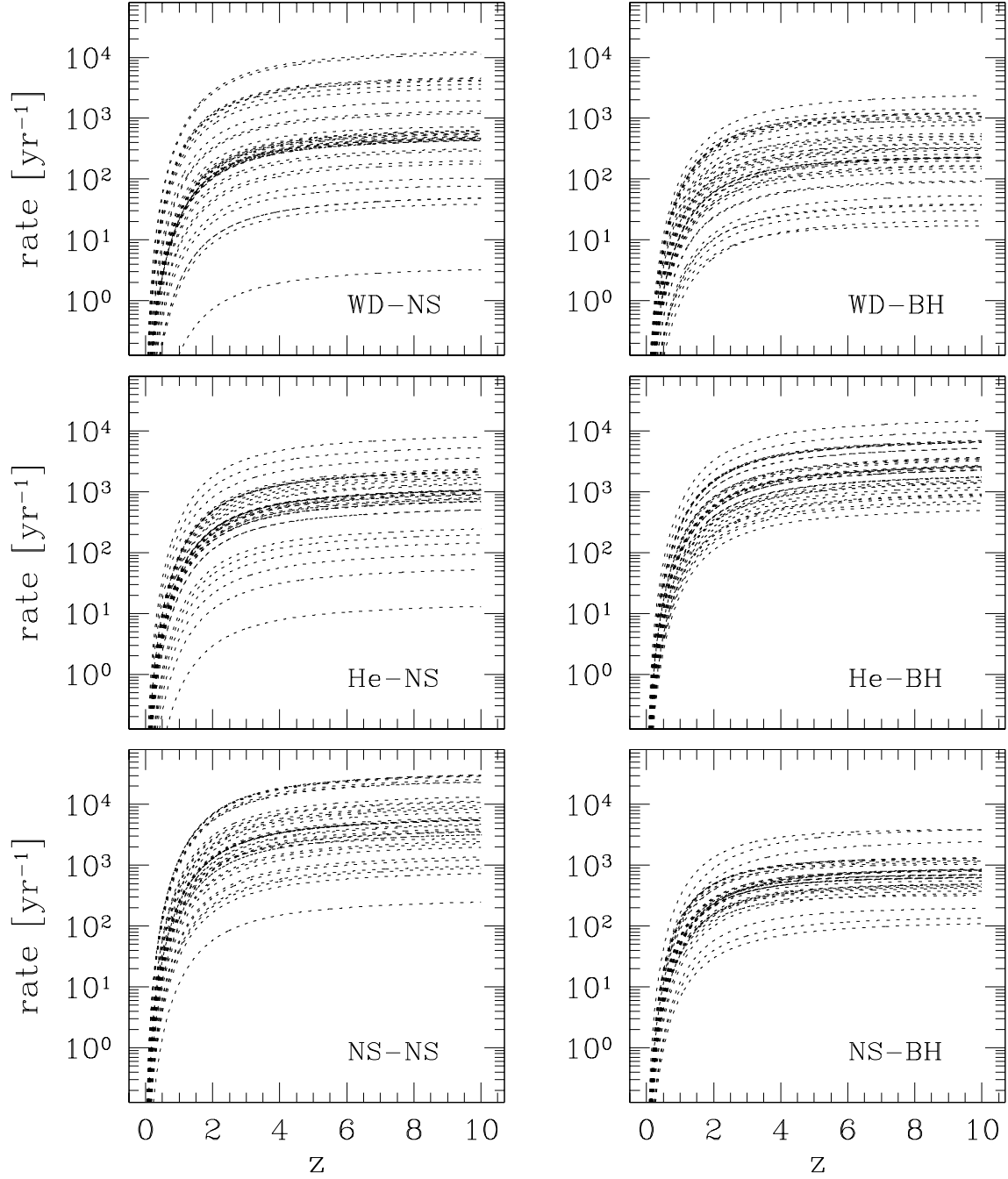,width=0.9\textwidth}}
\caption{
Cumulative event rates of different GRB progenitor types as a 
function of redshift for all our different models.
All rates were calculated with assumed star formation rate 
history of Rowan-Robinson (1999).
For all calculations flat cosmology model was used, with
$\Omega_{\rm m}=0.3$ and $\Omega_{\Lambda}=0.7$.
}\label{cosmoPS}
\end{figure*}

{\em Parameter Study}
The redshift dependence of GRB cosmic rate is presented for all the
different evolutionary models listed in Table~\ref{models}
  and for Rowan-Robinson (1999) SFR
in Figure~\ref{cosmoPS}.
For each progenitor type, we see that there are models which fail to
reproduce and at the same time there are always models which exceed,
sometimes significantly, the observed rate.
Moreover the rates for most of progenitors are very sensitive to the 
assumed evolutionary model.
Therefore, due to the population synthesis uncertainties, we are not
able to confirm  or reject any binary GRB progenitors just purely on
the basis of their rates. A similar note of caution should 
be added to any conclusions about GRB collimation based on the 
population synthesis results (e.g. Lipunov et~al. 1995, 
Portegies-Zwart and Yungelson 1998). The intrinsic spread in the rates when 
considering different population synthesis models is up to 
two orders of magnitude which corresponds to a factor of ten
in the estimates for collimation.

\subsection{Distribution around host galaxies}

{\em Standard Model}
In the  standard model (model A) we have evolved
$N_{\rm tot}=3 \times 10^7$ initial binaries and 4577 WD-NS,
2369 WD-BH, 9656 He-NS, 23494 He-BH, 52599 NS-NS and 8105 NS-BH
coalescing systems formed.
Next, we distributed the systems in a galactic disk and assigned
galactic velocities and propagated until the merger times
as described in \S\,2.2.
Besides the GRB progenitor systems, for each galaxy mass we also
propagated a number of coalescing WD-WD binaries, to trace the
 galactic distribution of stars.

\begin{figure*}[t]
\centerline{\psfig{file=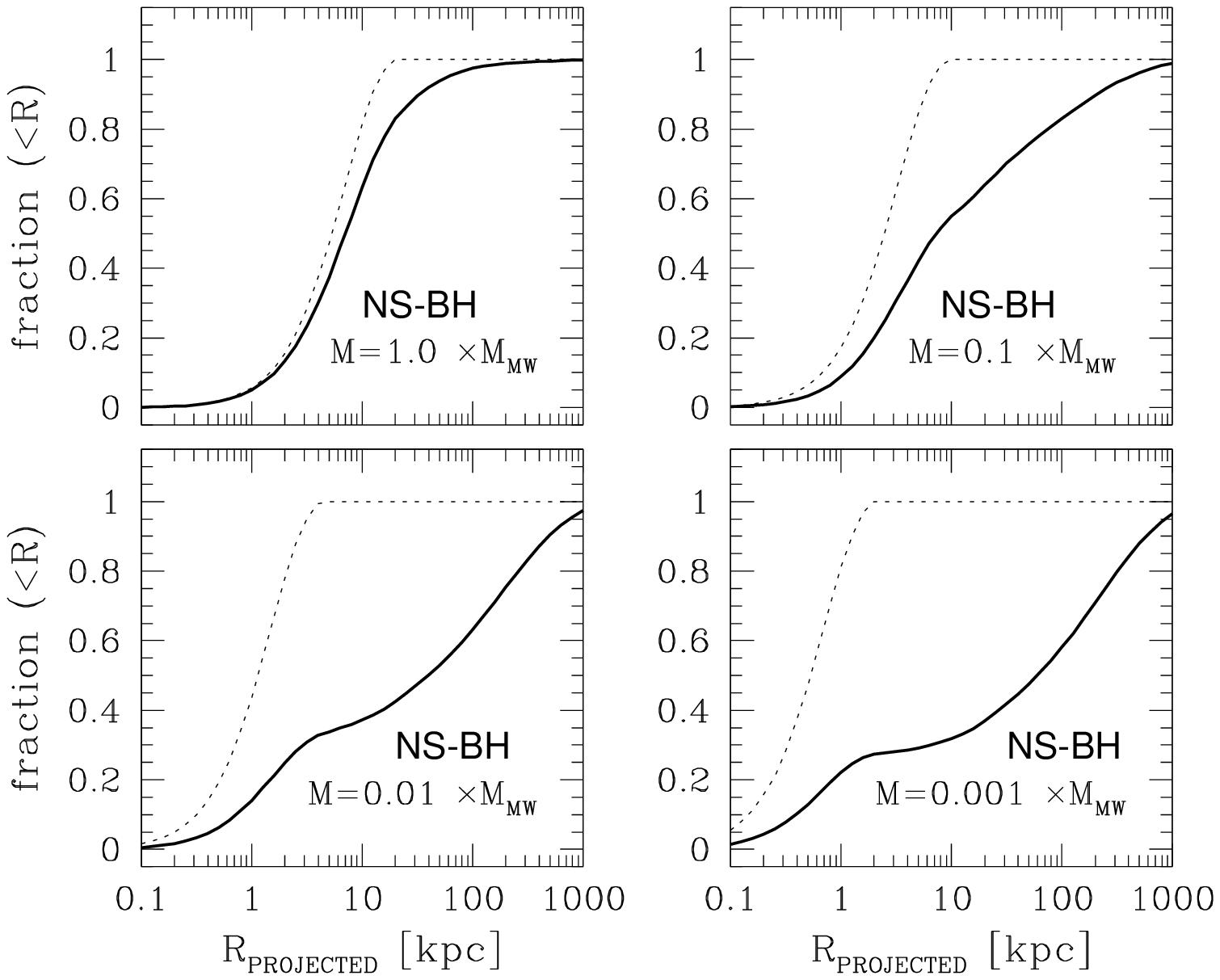,width=0.9\textwidth}}
\caption{Cumulative distributions of neutron star black hole
binaries merger sites around different mass galaxies (solid line)
for our standard evolutionary scenario  (model A).
The initial distribution of primordial 
binary population within the galaxy is shown with the dashed line.
}
\label{NSBH}
\end{figure*}

\begin{figure*}[t]
\centerline{\psfig{file=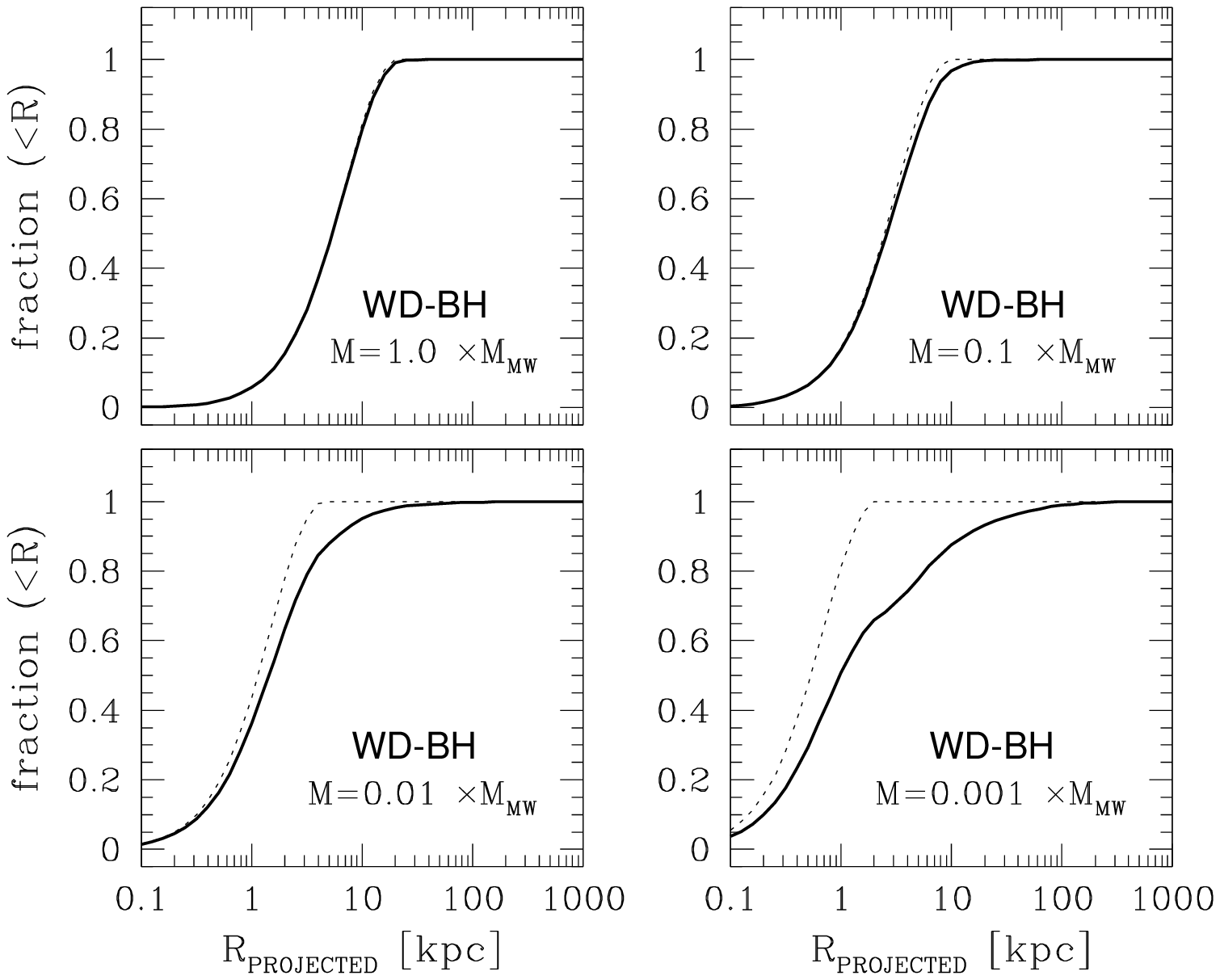,width=0.9\textwidth}}
\caption{Cumulative distributions of white dwarf black hole
binaries merger sites around different mass galaxies (solid line)
for our standard evolutionary scenario  (model A).
The initial distribution of primordial binary population within 
the galaxy is shown with the dashed line.
}
\label{WDBH}
\end{figure*}

\begin{figure*}[t]
\centerline{\psfig{file=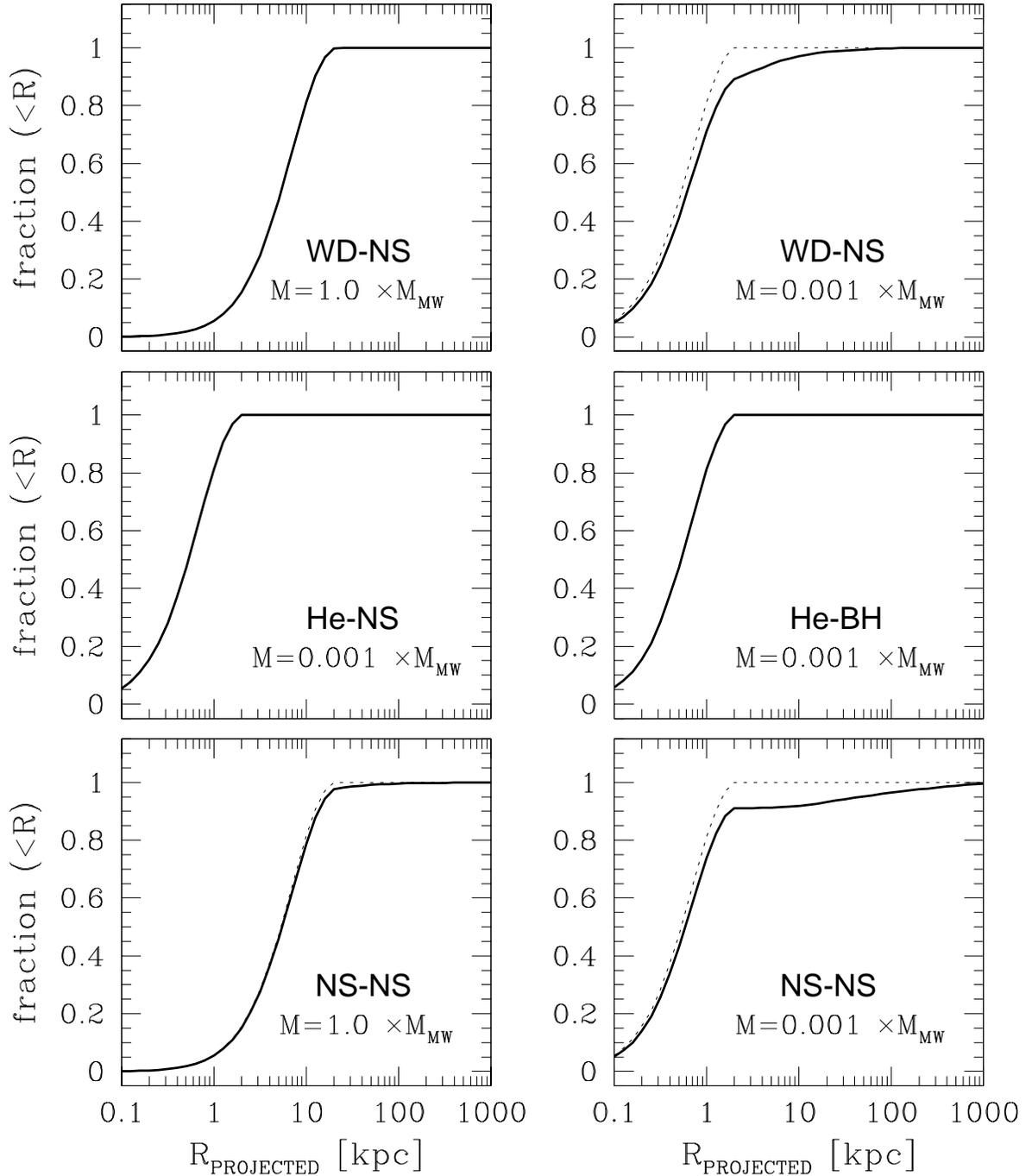,width=0.9\textwidth}}
\caption{Cumulative distributions of several types
of GRB progenitors around galaxies for our standard evolutionary 
scenario (model A). 
The case of white dwarf neutron star mergers is illustrated
in the top  panel with to extreme cases:
a Milky Way like galaxy (Left), and small galaxy with the mass 
$0.001 M_{MW}$.
We present the distributions of Helium star mergers in the middle 
panel for the case of a small galaxy only.
The lower panel contains the plots with the distributions of 
double neutron star mergers around a Milky Way like galaxy (Left), 
and around a small dwarf galaxy with the mass $0.001 M_{MW}$.
The initial distribution of primordial binary population within
the given galaxy is shown with the dashed line.
}
\label{NS}
\end{figure*}

We show the results of the propagation calculations in
Figures~\ref{NSBH},~\ref{WDBH},~\ref{NS}.
In each figure we show the cumulative distributions of the projected distances
of a given type merger.
The projected distance is the distance in the direction perpendicular to the
line of sight and we have averaged over all possible orientations
of the  host galaxy.
In each figure, we also show, with a dashed line, the initial
 stellar distribution within a galaxy of a given mass.
Note the different cut off radius ($R_{\rm max}$) of the initial
distribution at $20.0$, $9.3$, $4.3$ and $2.0$\,kpc for the 
four  galaxy masses defined in \S\,2.2.

The case of NS-BH mergers is shown in Figure~\ref{NSBH}.
The mergers spread out with decreasing mass of the galaxy and
even in the case of a large galaxy a significant number of
NS-BH mergers takes place outside of the host.
For a massive galaxy ($M_{MW}$) 20\% of NS-BH mergers will take place
outside the disk of the host, and as much as 70\% will escape hosts of 
small mass ($0.01-0.001 \times M_{MW}$).
This is due to the kicks that lead to velocities above the host
escape velocity and relatively long lifetimes of NS-BH binaries.

Mergers of WD-BH binaries take place within massive hosts while
a significant fraction escapes from low mass galaxies
(see Figure~\ref{WDBH}).
For galaxy masses $1-0.1 \times M_{MW}$ almost all WD-BH system
mergers trace their initial distribution.
For galaxy masses of $0.01, 0.001 \times M_{MW}$, 15\% and 35\%
WD-BH mergers take place outside of hosts, respectively.
These systems receive at most one kick during the evolution,
and the gain of the velocity is not large enough for these
binaries to escape from the potential well of a massive galaxy.
On the other hand, WD-BH systems have rather long lifetimes, and
if the  potential well is not deep enough to keep them inside
the galaxy, they escape and merge far away from the galaxies, as in 
the case of small mass hosts.

Lighter WD-NS systems tend to merge within host galaxies, with
only a slight dependence on the host mass (see Figure~\ref{NS}).
For massive galaxies, all their mergers take place close to
the places they were born, while for smallest galaxies up to 10\%
merge outside but close to the host outer regions.
Since they are lighter than the WD-BH systems, and on average they
receive higher kicks, one could expect that their mergers
should be spread out more than these of WD-BH binaries.
The distribution of the merger sites for a given mass galaxy is
in general the result of two competing effects; (i) the magnitude
of kicks the systems of a given type receive and (ii) the systems
characteristic lifetimes.
These two effects are not independent;
the binary lifetimes become smaller with stronger kicks because then 
only the tight, strongly bound systems survive.
As it turns out, for WD-NS systems the short lifetime effect dominates
over the velocity  effect, and although they receive higher kicks
they do not have enough time to travel outside the host before
the merger takes place.

Locations of He-NS and He-BH merger sites follow closely the 
initial distribution of their birth places, independently of 
the host galaxy mass (see Figure~\ref{NS}).
This is primarily due to their very short lifetimes but also
to their small systemic velocity gain.
Both, He-NS and He-BH systems have the shortest lifetimes of all
the  potential GRB progenitors studied here (see Table~\ref{Times}).
Their mergers take place even before the secondary finishes its
nuclear evolution (i.e., before it forms a remnant) in the CE phase,
when the secondary evolves off the main sequence and expands to
giant size.
For all the other progenitor types, both stars have to first form
the stellar remnants, and then usually considerable time is needed
for gravitational radiation to bring the two remnants together to a
final merger.
Also He-NS and He-BH systems are relatively heavy, so the one kick
the system experiences, does not have a great effect on the systemic
velocity.

Distribution of the projected distances of NS-NS merger sites follows
very closely the initial distribution of primordial binaries
(see Figure~\ref{NS}).
Only 2\% of NS-NS stars merge outside a massive host, and as
little as 8\% escape and merge outside of the lightest dwarf
galaxies.
These systems receive two  kicks, however due to their
very short merger times of the order $\sim 1$\,Myr  (see Belczynski
et al.\ 2001b for a discussion of the  merger time distribution) they 
predominantly merge within even the smallest hosts.
The NS-NS merger site distribution is quite similar to that of
WD-NS systems.
However, for NS-NS merger distribution there is a tail extending
to large distances from the host for  small mass galaxies.
This tail is due, to these systems which were formed along
classical channels.
These NS-NS have much longer merger times (typically $1-10$~Gyr)
than the rest of systems formed through one of the newly
recognized pathways.
The small contribution of these systems to the entire population of
coalescing NS-NS binaries does not change the overall tendency of
NS-NS to merge within even the lowest mass hosts.

\begin{figure*}[t]
\centerline{\psfig{file=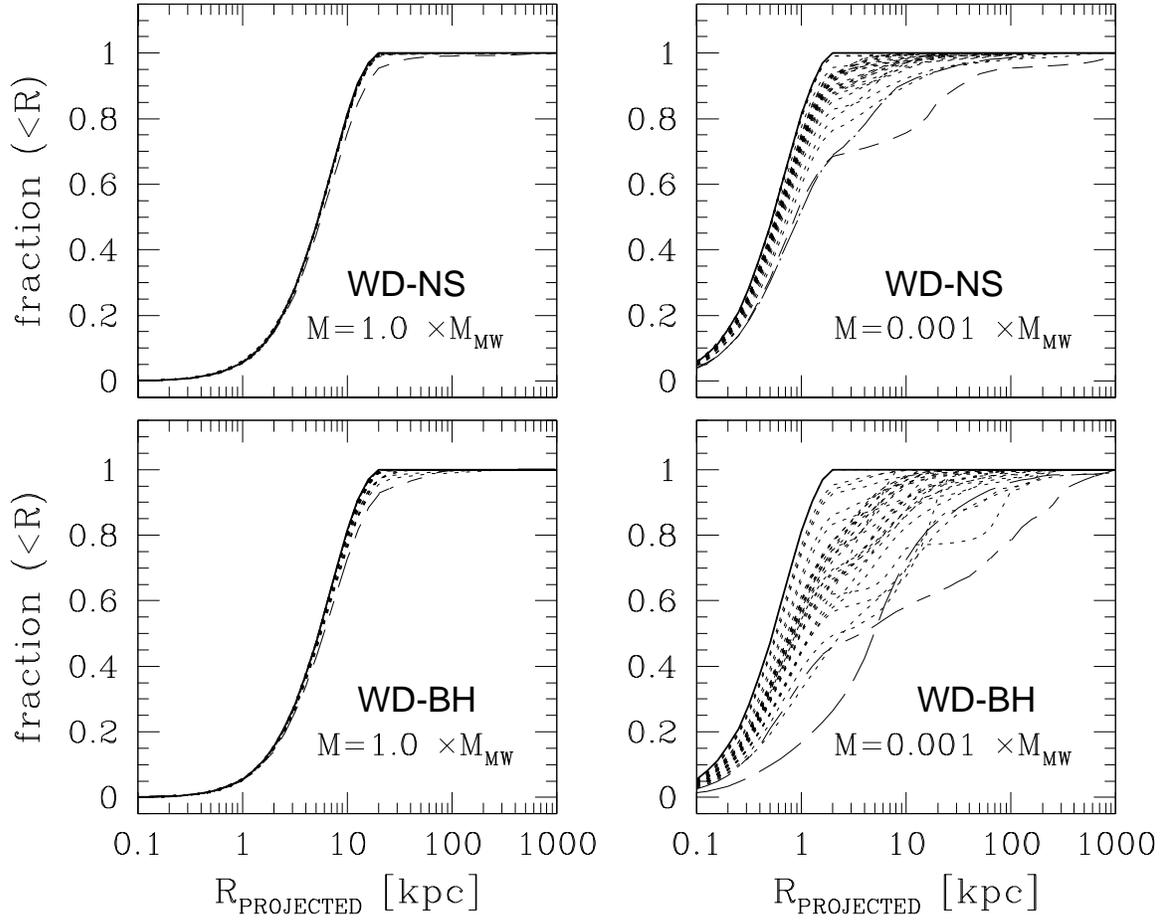,width=0.9\textwidth}}
\caption{
Cumulative distributions of WD-NS and WD-BH merger sites for two 
extreme galaxy masses and for different evolutionary models.
All models are shown with the dotted lines, except the most extreme 
ones: model F2 -- short dashed line, N -- doted long dashed line, 
E3 -- short long dashed line, L1 -- long dashed line. 
The initial distribution of primordial binary population within
the galaxy is shown with the solid line.
}
\label{WDPS}
\end{figure*}

\begin{figure*}[t]
\centerline{\psfig{file=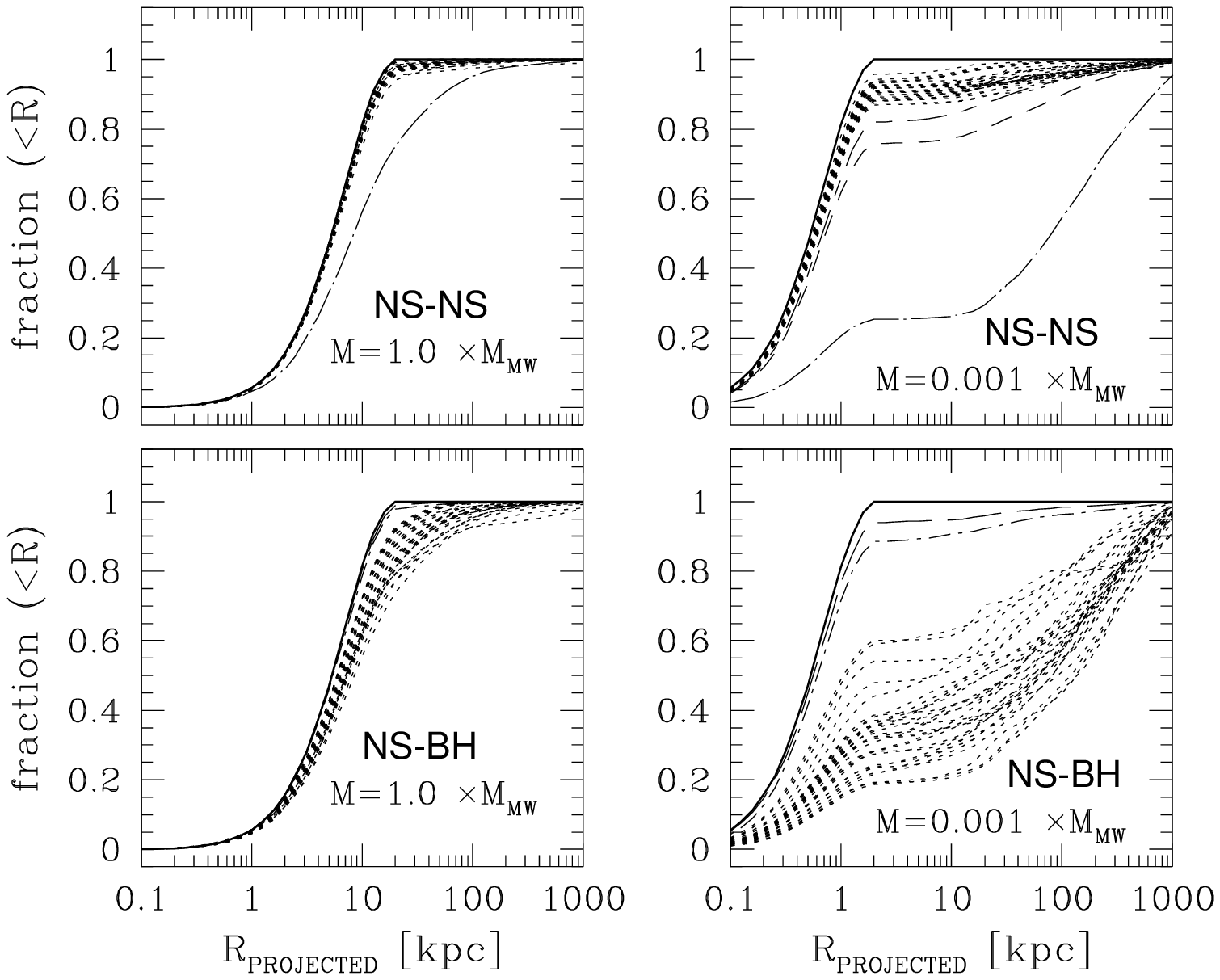,width=0.9\textwidth}}
\caption{
Cumulative distributions of NS-NS and NS-BH merger sites for two 
extreme galaxy masses and for different evolutionary models.
All models are shown with the dotted lines, except the most extreme 
ones: model F2 -- short dashed line, N -- doted long dashed line,
E3 -- short long dashed line, D1 -- dotted short dashed line, 
D2 -- long dashed line. 
The initial distribution of primordial binary population within 
the galaxy is shown with the solid line.
}
\label{NSPS}
\end{figure*}

{\em Parameter Study}
To study the dependence of our results on the assumed evolutionary
parameters and initial distributions, we have calculated distributions
of GRB progenitor merger sites for all models listed in Table~\ref{models}.
For all our models, and for all simulated galaxy masses,
He-NS and He-BH mergers follow  the initial distribution of
initial binaries, and merge within their host galaxies.
For all other systems, the results of our calculations are presented
in Figure~\ref{WDPS},~\ref{NSPS}, for two extreme host galaxy masses of
$1.0\ M_{MW}$ and  $0.001 \times M_{MW}$.

The distribution of merger sites of WD-NS systems is rather independent
of the model parameters and these systems merge mainly within host
galaxies, irrespective of the host mass.
Most of the models are concentrated around the standard model
distribution.
Just in a few models more than 10-15\% of WD-NS mergers take place
outside of the smallest hosts.
The two most extreme cases were identified in Figure~\ref{WDPS} and
they correspond to models N and F2.
Model N represents nonphysical case of stellar evolution (and shall be
treated as such), in which no helium giant radial evolution
is allowed.
This model was calculated just for comparison with previous results
which did not take in to account  this effect.
Model F2, represents evolution in which every mass transfer episode 
(except CE phase) is treated conservatively, i.e., all material lost from
the donor is accreted by the companion ($f_a=1$).
The effect of such a treatment, as compared to our standard evolution
where half of the material is lost form the system, is that 
post-MT systems have wider separations, since no material and thus no
angular momentum is lost from the binary.
Naturally, the final WD-NS binaries are wider as well, and have longer
merger times, which allow some systems to escape from host galaxies.
Such a model is  rather extreme, as we know
that during MT events material is lost from at least some
systems (e.g., Meurs \& van den Heuvel 1989).

Distributions for WD-BH merger sites show quite significant
spread, allowing the possibility that majority of these systems
merge outside the low-mass hosts.
Although for massive hosts, most of the models show that these
systems merge within the host boundary, for low mass galaxies as much
as 40\%, or even more merge outside the hosts.
Two most extreme cases are these for models designated as E3 and L1.
For both models, the systems formed after CE or a MT phase are
wider than for our standard evolutionary scenario.
Therefore, it is natural that WD-BH binaries have longer merger
times and have greater chances of escaping the hosts.
As we double the CE efficiency to $\alpha_{\rm CE} \times
\lambda =2$) in model E3, during the CE phase binaries use much less 
of their orbital energy to expel the common envelope.
Due to this smaller energy loss post-CE binaries are left  with wider
orbital separations than they would have for $\alpha_{\rm CE} \times
\lambda =1$ of our standard model.
Decreased to half of its value the angular momentum loss, $j=0.5$
of model L1, directly influences the separations of post-MT systems.
And although, in this model some material is lost from the systems
during MT, unlike for model F2 discussed above, the angular momentum
loss is much decreased, so binaries are much wider than for our
standard model ($j=1$).

Mergers of NS-NS predominantly take place inside host galaxies.
For massive host, all models follow very closely the initial
distribution  binaries, and depending on the model
95\% or more of mergers take place within
massive hosts.
For lowest mass galaxies,  all but two models, give $\sim$ 90\%
or more mergers within a host boundary of 2 kpc.
Models E3 and F2 stand out, but still even for
these two, more than $\sim$ 80\% of NS-NS mergers happen within
the smallest mass hosts.
In the above, we haven't taken into account model N, marked in
Figure~\ref{NSPS} with a dot long dashed line.
As mentioned before, this is an nonphysical model, and is shown here
just for comparison with previous results.
In agreement with previous calculations (e.g., Belczynski
et al.\ 2000) for the case of massive galaxy, in model N about
30--50\% of NS-NS stars merge outside host, or further away from
host than 10--20 kpc.
For lowest mass galaxies, in model N, as many as $\sim$ 70\% of NS-NS
mergers take place outside hosts, or a few kpc from the center of the
hosts.
Detailed discussion of this significant change of results for
NS-NS is presented in Belczynski et al.\ (2001a).

Distribution of NS-BH merger sites around host galaxies is quite
sensitive to the model parameters.
Although for the case of propagation in the potential of a very massive
galaxy at least 70\% of these systems merge within hosts, majority
of these mergers takes place far away from low-mass hosts.
The curves corresponding to models D1 and D2 clearly differ from all the remaining
distributions.
Since these two models, have lower maximum NS mass
($M_{\rm max,NS}=2.0,1.5 M_\odot$), many objects classified in the 
standard model ($M_{\rm max,NS}=3.0 M_\odot$) as NS-NS are included
as NS-BH in the distributions of models D1 and D2.
These is the reason why the NS-BH distributions in models
D1 and D2 resemble the standard model
NS-NS distribution.
If in fact, the maximum neutron star mass is much lower than our
assumed $3.0 M_\odot$, most of the NS-BH are expected to merge
within even small galaxies, with only the heaviest binaries escaping
their hosts.

\subsection{Comparison of the merger sites with GRB observations}

The discovery of gamma-ray burst afterglows by the Beppo SAX
satellite have lead to the identification of GRB host galaxies, and to
the localization of GRB events with respect to these galaxies.
In Table~\ref{Loc} we list the data on GRB
positions around host centers. Most of these are
taken from Bloom et al.\ (2001), and we have added
the entries for three recent bursts.

From Table~\ref{Loc} we see that GRBs take place not far from
the  centers of their host galaxies.
For some bursts, GRB970508, GRB000418 and GRB010222, the offsets are very
small, and positions of the optical afterglows are coincident with host centers.
Moreover, the host galaxies are typically small,  irregular, with intense
star formation (e.g., Fruchter et al.\ 1999; Holland 2001;
Ostlin et al.\ 2001; Bloom et al.\ 2001).
One has to note that the data presented in Table~\ref{Loc} describes only the
long GRBs since only for these bursts afterglows have been  observed so far.

Comparing the theoretical distributions like these
calculated above with observations is a difficult task.
Ideally one would like to compute the theoretical distribution
of angular offsets between the GRB and its nearest galaxy, taking into account
the fact that the nearest galaxy may not necessarily be the host galaxy.
Such a calculation would require  a number of assumptions,
about the evolution of galaxies with redshift, about the
rate of star formation in galaxies, and about the mass and size
distribution of galaxies as a function of redshift.
Each of these quantities is uncertain by itself.
Thus a calculation like that in our current knowledge
of the evolution of galaxies would depend on a number
of uncertain assumptions and thus could lead to very uncertain
results. 
However, results of calculations, which take into account some of the 
listed above effects, were independently obtained and presented by Perna 
\& Belczynski (2001).
Here we adopt a more straightforward approach.
We assume a cosmological model with $H_0=65$ km\, s$^{-1}$\,Mpc$^{-1}$,
$\Omega_M=0.3$, and $\Omega_\Lambda=0.7$, and calculate the
physical distance to the galaxy claimed to be  the host galaxy.
We then use the Kolmogorov-Smirnov test (e.g., Press et al.\ 1992)
to verify the hypothesis that the observed distribution of offsets
has been drawn from the distribution of offsets for a given type
of GRB progenitor around a galaxy of a given mass.
In each case we repeat such calculation for a number of population synthesis models
listed in Table~\ref{models} to assess the range
of systematic errors introduced by the population synthesis.

We present the results of these calculations in Table~\ref{KS}.
For each progenitor type we list the Kolmogorov-Smirnov test probabilities
in case of the four galaxy masses defined in \S\,2.2.
We also list the   highest and the lowest probability obtained when different
models (with the exception of nonphysical model N) of population synthesis were used.
Table~\ref{KS} allows to evaluate the viability
of each type of the GRB progenitor.

Let us assume in this discussion that we reject a given hypothesis
if the KS test probability is below 1\%.
One thing becomes immediately  clear from Table~\ref{KS},
i.e. GRB progenitors do not reside in large galaxies like
the Milky Way. GRB afterglows are related to small galaxies with
the masses around  $0.01 M_{MW}$ ($0.015 \times 10^{11} M_\odot$).
This has been noted by the observers claiming that the typical host galaxy
mass will lie in the range of $0.001-0.1 \times 10^{11} M_\odot$
(e.g., Ostlin et al.\ 2001, Bloom et al.\ 2001).
It has to be noted that for each type of progenitor
the value of the Kolmogorov-Smirnov test probabilities
is a strong function of the model galaxy used.

Double neutron star mergers are an acceptable choice
and in the case of a low mass galaxy   $0.01 M_{MW}$,
the probability that the observed offset distribution is
the same as the theoretical one is very high.
Thus inclusion of the additional formation channels for this
type of binaries has a significant effect (e.g., this possibility
was rejected by Bloom et al.\ 2001).
The range of probabilities covered by several different models
of population synthesis is small. 
In our models the population of NS-NS mergers is dominated by short
lived systems. Only in the case of a very low mass galaxy
$0.001 M_{MW}$, do the kick velocities play a role.
Here the lowest KS test probability  corresponds to the model B1
with very small kick velocities B1.
Since, for the smallest kicks, NS-NS binaries form in 
wider orbits and with longer lifetimes (e.g., Kalogera 1996) they have
more time to escape from their host galaxies.

The Kolmogorov-Smirnov test results presented in Table~\ref{KS}
show that we can certainly reject 
NS-BH mergers as GRB progenitors. The highest probability
is obtained for the case of a $0.1 M_{MW}$ mass galaxy, but its value
is still not acceptable.
This is due to the fact that these binaries are rather long
lived (see Table~\ref{Times}), therefore NS-BH can escape from host
galaxies and merge far away from host centers.
KS test probabilities
rise to acceptable values in models D1 and D2, where the maximum mass
of a neutron star is lower. In such models a number of
binaries classified typically  as double neutron stars
contribute to the NS-BH population.

The values of the probabilities for the mergers involving white dwarfs
(WD-NS and WD-BH) are large and make these models acceptable.
In the case of WD-BH mergers the probability even rises to 0.94 for the case
of a very small mass $0.001 M_{MW}$  galaxy, however
this number is rather uncertain due to the very wide range of K-S probabilities
obtained for different population synthesis models. These uncertainties
are not that large in the case of $0.01 M_{MW}$ and larger galaxies.

Similarly we can not reject the He-BH and He-NS mergers.
In these cases the probabilities are not as large as in the case of WD
mergers, however these groups
also constitutes a viable GRB  progenitor.  One should note
that the systematic errors due to different population synthesis
models are very small in this case, and the main factor that influences
the value of the KS test probability is the distribution
of stars in a model galaxy. 
He-BH and He-NS mergers evolve on  very short timescales and therefore 
take place in star forming regions.

\subsection{Comparison with other studies}

{\em Merger Rates}
A comparison of our merger rates of NS-NS and NS-BH binaries with
a number of other studies has been discussed in detail by Belczynski
et al.\ (2000b).
In short, our rates are in good  agreement with previous
theoretical predictions.
Although we have noted some significant differences, we attributed
them to the more approximate treatment of stellar and binary
evolution in earlier studies and to our recognition of new NS-NS
populations.
Recognition which was based on the assumption that CE phases initiated 
by evolved low-mass helium stars do not always lead to binary component 
mergers.

Merger rates for several other binary GRB candidates have been so far
presented only by Fryer et al.\ (1999a).
We are not able to directly compare the rates because Fryer et al.\
(1999a) did not define the masses of WD selected to enter the WD-BH GRB
progenitor candidate group.
We encountered a similar problem in the case  of He-BH mergers, for which 
the masses of He cores are not given.
However, if we  assume that their systems correspond to our
definition of GRB binary candidates, then we note a very close
resemblance of the He-BH the rates, and a rather good agreement of WD-BH
rates.
Fryer et al. (1999a) found their rates of $0.15\,$Myr$^{-1}$ for WD-BH 
and $14\,$Myr$^{-1}$ for He-BH
for their standard evolutionary model.
In our models closely resembling the Fryer et al.\ (1999a) standard model, 
in particular these with smaller binary fraction (K1) and higher kicks (B9-12), 
we predict WD-BH rates of $0.4-1.0\,$Myr$^{-1}$ and He-BH rates of 
$5.8-12.8\,$Myr$^{-1}$.

{\em Host Merger Site Distributions}
We may compare our results to those of Bloom et al. (1999), 
Bulik et al. (1999), Fryer et al.\ (1999a), and Belczynski 
et al. (2000).
These authors calculated distributions of NS-NS and NS-BH
mergers around different mass galaxies.
The main conclusion of these studies was that a significant
fraction (up to 40\% for massive hosts, and up to 80\% for low
mass hosts) of NS-NS and NS-BH binaries merge outside of host
galaxies.
For NS-BH binaries we find very good agreement with previous
studies, as we find that up to 25\% and 80\% of these binaries
will merge outside a massive and low mass host
(see Figure~\ref{NSPS}).
Although our calculations show bigger concentration of NS-BH
mergers in massive hosts, it is explained by the fact that we
have adopted decreased kicks for BH, and therefore systemic
velocity gain is decreased as well.
However, our conclusions  for NS-NS mergers are very different
from all previous studies, due to the newly recognized
short lived populations of these binaries, as discussed
throughout this work.

Both Fryer et al.\ (1999a) and Bloom et al.\ (2001) assumed
that He-BH mergers will take place in the star formation
regions of the  host galaxies.
With our calculations we may confirm that, in fact, He-BH and
He-NS merger sites, follow exactly star formation regions in
their host galaxies.

Fryer et al.\ (1999a) argued that WD-BH merger sites are also
concentrated within host galaxies, a conclusion that was later
adopted by Bloom et al.\ (2001).
Our detailed calculations show that, in fact, for massive
galaxies these systems follow closely the initial
primordial binary  distribution, and merge within hosts.
However, for small mass galaxies, a significant fraction of WD-BH
binaries merge outside of hosts.
Depending on the assumed evolutionary model as many as 50\% of
these systems may merge outside of small mass hosts
(see Figure~\ref{WDPS}) as discussed in \S\,3.6.
Fryer et al.\ (1999a) argued that as these systems have very short
merger times of $\sim 100$\,Myr, and they will not have enough time
to escape from hosts.
We find that the merger times of these systems are indeed
of order of $\sim 100$\,Myr (see Table~\ref{Times}).
Nevertheless, the actual calculations of WD-BH trajectories
prove that this conclusion  is not valid for small mass and size
hosts ($0.01-0.001 \times M_{\rm MW}$).

\section{Discussion and conclusions}

We have presented calculation of rates and  spatial
distributions around host galaxies
of several binary merger events, which were proposed
as possible GRB progenitors.
We have used the {\em StarTrack} population synthesis
code in our calculations.

We have found that the  rates are very sensitive to the assumed set of
 stellar evolutionary parameters.
Using the rates alone we were not able to
exclude any of the proposed  binaries as GRB progenitors, since
the highest rates obtained were
always higher than observed BATSE GRB rate for any type of a binary.
In the framework of the standard population synthesis model (model A)
we find that the total rate of all the proposed binary
events is roughly ten times larger than the observed GRB rate.
However, we find that the spread in the rates due to uncertainties in population
synthesis is large, and in some cases exceeds 
a factor of $\sim 100$. This corresponds to the uncertainty in the
estimate of the collimation of a factor of ten.
On the other hand we note that our standard model (model A) 
leads to an expected collimation half--opening angle
of $\Theta \ga 25^\circ$.
The measured collimation angles are somewhat smaller than this
value, typically a few degrees (Panaitescu \& Kumar 2001).
Estimates of the GRB rates or collimation based 
on population synthesis alone carry  large systematic    
errors.

Distributions of binary system merger sites around galaxies
 may be compared
to the locations of GRB optical afterglows with respect
to the galaxies identified as their  hosts.
Most of GRBs take place inside or close to the host galaxies 
 (e.g., Bloom et al.\ 2001).
Observed GRB hosts are  small-mass galaxies, often
thought to be going through vigorous star formation phase.

There are no reliable GRB host mass estimates, and thus 
we have calculated models for a range of galaxy masses.
Our standard model calculations were repeated for a number
of different evolutionary models to assess the robustness of
our results.
We have found that the NS-BH mergers  take place
mainly outside of their host galaxies, and thus are inconsistent 
with the observed locations of GRBs around hosts. 
Some WD-BH binaries may merge outside the star formation regions
of their host galaxies. 
However, the distribution of the WD-BH merger sites  around their 
host galaxies is consistent with the observed distribution of
GRB offsets from the centers of galaxies.
Thus one can not reject the WD-BH mergers purely
on the basis of comparison with the observed offsets.
However, if one additionally requires that the mergers should
take place in the proximity of star forming regions in galaxies then
the WD-BH mergers  can be rejected as potential GRB candidates.
Merger sites of WD-NS, He-NS, He-BH, and NS-NS trace
the star formation regions of the hosts,
for all  the cases of the host mass and size considered here,
and independently  of  the
adopted population synthesis model.
We conclude that these types of binaries may be responsible at least
for a part of observed GRBs.

GRBs form a very  nonuniform group of events, with different outburst
times, very different light curves and observed energies.
Thus, there is a possibility that GRBs originate in more than 
one type of progenitor.
Locations of GRBs with respect to host
galaxies  has  so far been measured only for long GRBs.
There is a growing evidence that these GRBs are related to
collapsing  massive stars: collapsars.
However,  our results show that  several types of binary
system progenitors cannot be rejected purely on the basis of
their merger site distribution.
Additionally, if binaries were responsible for only a part of the observed
GRBs, we also can not exclude them purely on the basis of their
expected coalescence rates.

Because of the expected short duration times, NS-NS and NS-BH 
mergers are the primary candidates as short burst progenitors.
These two populations exhibit  very different distributions
of merger sites. 
Mergers of NS-NS systems take place predominantly within hosts, 
to the contrary of what was so far believed, provided that CE phases
initiated by low-mass helium stars do not always lead to binary 
component mergers, assumption which yet have to be tested by detailed
hydrodynamical calculations. 
On the other hand, a significant fraction of NS-BH systems merge 
outside of their host galaxies.
At some point in the future afterglows from short GRBs will be 
observed and their locations with respect to host galaxies will be 
measured, and then such calculations may provide a useful tool to 
distinguish between these two progenitor models 
(Perna \& Belczynski 2001).
Future and current space missions like HETE-II, INTEGRAL,
GLAST or SWIFT will hopefully measure precise
positions of a large number of bursts even of the short
duration and  settle down the issue of GRB progenitors.

\acknowledgements
We are indebted to several people for very useful discussions on the
various aspects of this project.
In particular we want to thank Vicky Kalogera, Chris Stanek,
Rosalba Perna, Stephen Holland and Boud Roukema.
Support is acknowledged by the Polish Nat.\ Res.\ Comm. (KBN)
grant 5P03D01120 to KB and TB.
KB acknowledges support from the Smithsonian Institution through a  
Predoctoral Fellowship, from the Lindheimer fund at Northwestern 
University, from the Polish Science Foundation (FNP) through a 2001
Polish Young Scientist Award

\pagebreak

\begin{deluxetable}{crcrc}
\tablewidth{350pt}
\tablecaption{ Characteristic Timescales of GRBs Candidates (Myr)}
\tablehead{ Type& $t_{\rm evol}$\tablenotemark{a}& $\Delta t_{\rm evol}$& 
$t_{\rm merg}$\tablenotemark{b}& $\Delta t_{\rm merg}$ } 
\startdata
WD-NS& 26.6& 20.0--35.7&    6.8& 0.014--1238 \\
WD-BH& 25.6& 20.1--34.6&   96.9& 16.1--1831 \\
He-NS& 10.0& 7.04--11.1&    ...& ... \\
He-BH&  7.9& 5.62--9.96&    ...& ... \\
NS-NS& 18.5& 10.7--27.8&    0.7& 0.017--390 \\
NS-BH&  7.7& 5.92--17.4&  534.6& 1.68--5170 \\

\enddata
\label{Times}
\tablenotetext{a}{distribution median of evolutionary time delay}
\tablenotetext{b}{distribution median of merger time delay}
\end{deluxetable}

\begin{deluxetable}{ll}
\tablewidth{350pt}
\tablecaption{Population Synthesis Model Assumptions}
\tablehead{ Model & Description}
\startdata

A      & standard model described in \S\,2 \\
B1--13 & zero kicks, single Maxwellian with \\
       & $\sigma=10,20,30,40,50,100,200,300,400,500,600$\,km\,s$^{-1}$, \\
       & ``Paczynski'' kicks with $\sigma=600$\,km\,s$^{-1}$ \\
C      & no hyper--critical accretion onto NS/BH in CEs \\
D1--2  & maximum NS mass: $M_{\rm max,NS}=2, 1.5$\,M$_\odot$ \\
E1--3  & $\alpha_{\rm CE}\times\lambda = 0.1, 0.5, 2$ \\   
F1--2  & mass fraction accreted: f$_{\rm a}=0.1, 1$ \\
G1--2  & wind changed by\ $f_{\rm wind}=0.5, 2$ \\
H      & Convective Helium giants: $M_{\rm conv}=4.0$$\,M_\odot$ \\ 
I      & burst--like star formation history \\
J      & primary mass: $\propto M_1^{-2.35}$ \\
K1--2  & binary fraction: $f_{\rm bi}=0.25, 075$ \\ 
L1--2  & angular momentum of material lost in MT: $j=0.5, 2.0$\\
M1--2  & initial mass ratio distribution: $\Phi(q) \propto q^{-2.7}, q^{3}$\\
N      & no helium giant radial evolution\\
O      & partial fall back for $5.0 < M_{\rm CO} < 14.0 \,M_\odot$\\ 
P1--2  & minimum Helium core mass in He-NS/BH mergers $M_{\rm min,He}= 4, 8 \,M_\odot$\\
R1--2  & minimum WD mass in WD-NS/BH mergers: $M_{\rm min,WD}=0.7, 1.1 \,M_\odot$ \\

\enddata
\label{models}
\end{deluxetable}

\begin{deluxetable}{crrrrrr}
\tablewidth{350pt}
\tablecaption{ Galactic Binary GRB Progenitors Coalescence Rates (Myr$^{-1}$)}
\tablehead{ Model\tablenotemark{a}& WD-NS& WD-BH& He-NS& He-BH& NS-NS& NS-BH } 
\startdata

A    &   4.6&  2.4&  9.7& 23.5&  52.7&  8.1 \\
B1   &  46.3& 13.4& 20.9& 64.2& 292.4& 18.2 \\
B2   &  50.9& 12.9& 21.4& 62.9& 299.6& 19.4 \\ 
B3   &  48.7& 13.6& 20.8& 63.7& 302.2& 19.6 \\ 
B4   &  44.6& 12.2& 20.7& 66.5& 285.2& 19.1 \\
B5   &  38.2& 11.3& 22.8& 67.2& 251.0& 19.5 \\
B6   &  32.2& 10.3& 19.9& 64.0& 226.8& 16.4 \\
B7   &  13.4&  5.3& 15.2& 48.9& 128.1& 14.6 \\
B8   &   4.8&  2.6&  9.9& 23.5&  57.5& 10.1 \\
B9   &   1.9&  0.9&  8.9& 12.8&  33.2&  5.7 \\
B10  &   0.8&  0.9&  6.9&  9.7&  18.2&  3.7 \\
B11  &   0.4&  0.4&  6.2&  7.6&  12.5&  2.1 \\
B12  &   0.4&  0.4&  4.6&  5.8&   8.2&  1.5 \\
B13  &  12.2&  4.0& 12.3& 29.8&  91.0& 10.3 \\
C    &   0.4&  1.7& 33.3& 12.7&  43.2&  5.6 \\
D1   & 104.8&  7.7&  1.8& 33.8&  33.6& 23.3 \\
D2   & 114.7& 22.1&  0.1& 32.4&   9.1& 36.2 \\
E1   &  0.03\hspace*{-1.7mm}&  0.2&  0.5& 91.6&   2.5&  4.7 \\
E2   &   1.7&  0.3&  8.5& 47.8&  23.5&  6.3 \\
E3   &   5.4&  6.0&  4.6&  8.1& 109.0&  8.7 \\
F1   &   4.2&  2.1&  2.3& 14.5&  22.1&  9.3 \\
F2   &   6.5& 11.1&  8.2&  4.5&  54.3&  8.6 \\
G1   &   5.7&  5.8&  7.2& 20.3&  43.9& 14.2 \\
G2   &   4.8&  0.6& 19.7& 15.1&  94.8&  1.3 \\
H    &   4.7&  2.0&  8.2& 24.3&  37.9&  7.8 \\
I    &   4.3&  3.5&  9.7& 23.9&  54.5& 10.0 \\
J    &   4.8&  3.8& 12.6& 34.8&  58.1& 12.8 \\
K1   &   1.9&  1.0&  4.1&  9.9&  22.5&  3.4 \\
K2   &   7.8&  4.0& 16.3& 39.6&  90.2& 13.5 \\
L1   &   6.0&  3.6&  6.9&  8.4&  78.9&  9.2 \\
L2   &   4.3&  2.2&  6.6& 21.3&  12.0&  6.2 \\
M1   &   0.9&  4.3&  1.2&  5.9&   6.2&  4.0 \\
M2   &   5.8&  0.2& 17.4& 22.9& 114.2&  8.4 \\
N    &   7.5&  4.0&  8.7& 22.3&  34.4& 10.7 \\
O    &   4.2&  1.6& 10.0& 25.0&  51.9&  5.7 \\
P1   &   4.6&  2.4& 73.6& 33.3&  52.7&  8.1 \\
P2   &   4.6&  2.4&  0.9& 10.7&  52.7&  8.1 \\
R1   &   5.6&  3.5&  9.7& 23.5&  52.7&  8.1 \\
R2   &   2.9&  1.4&  9.7& 23.5&  52.7&  8.1 \\

\enddata
\label{rates}
\tablenotetext{a}{for definition of models see Table~\ref{models}}
\end{deluxetable}

\begin{deluxetable}{lllll}
\tablewidth{350pt}
\tablecaption{Location of GRB afterglows in relation to their host galaxies \tablenotemark{a}}
\tablehead{ GRB& redshift&  Offset $\Delta \Theta$ & R$_{\rm PROJECTED}$\ [kpc]& Comments } 
\startdata

970228  & 0.695     &  $0.426 \pm 0.034$''& $3.266 \pm 0.259$ & edge of host \\
970508  & 0.835     &  $0.011 \pm 0.011$''& $0.091 \pm 0.090$ & host center \\ 
970828  & 0.958     &  $0.474 \pm 0.507$''& $4.047 \pm 4.326$ & edge/outside \\
971214  & 3.418     &  $0.139 \pm 0.070$''& $1.105 \pm 0.557$ & inside host \\
980326  & $\sim 1$  &  $0.130 \pm 0.068$''& ...               & edge/outside? \\
980329  & $\la 3.5$ &  $0.037 \pm 0.049$''& ...               & inside host \\
980425  & 0.008     & $12.550 \pm 0.052$''& $2.337 \pm 0.010$ & inside host \\
980519  & ...       &  $1.101 \pm 0.100$''& ...               & inside host \\
980613  & 1.096     &  $0.089 \pm 0.076$''& $0.782 \pm 0.666$ & ??? \\
980703  & 0.966     &  $0.040 \pm 0.015$''& $0.038 \pm 0.128$ & inside host \tablenotemark{b} \\
981226  & ...       &  $0.749 \pm 0.328$''& ...               & ??? \\
990123  & 1.600     &  $0.669 \pm 0.003$''& $6.105 \pm 0.027$ & edge of host \\
990308  & ...       &  $1.042 \pm 0.357$''& ...               & ??? \\
990506  & 1.310     &  $0.297 \pm 0.459$''& $2.680 \pm 4.144$ & ??? \\
990510  & 1.619     &  $0.066 \pm 0.009$''& $0.600 \pm 0.084$ & edge of host \\
990705  & 0.840     &  $0.872 \pm 0.046$''& $7.165 \pm 0.380$ & inside host \\
990712  & 0.434     &  $0.049 \pm 0.080$''& $0.301 \pm 0.486$ & inside host \\
991208  & 0.706     &  $0.196 \pm 0.097$''& $1.513 \pm 0.750$ & edge?\\
991216  & 1.020     &  $0.359 \pm 0.032$''& $3.107 \pm 0.280$ & inside?\\
000301C & 2.030     &  $0.069 \pm 0.007$''& $0.622 \pm 0.063$ & inside? \\
000418  & 1.118     &  $0.023 \pm 0.064$''& $0.202 \pm 0.564$ & host center \\
000926  & 2.066     &  $1.5 \pm 0.5$''    & $13.43 \pm 4.5$   & edge/inside \tablenotemark{c} \\
010222  & 1.477     &  $0.05 \pm 0.05$''  & $0.45 \pm 0.45$   & inside host \tablenotemark{d} \\ 

\enddata
\label{Loc}
\tablenotetext{a}{all data from Bloom et al.\ (2001), but:}
\tablenotetext{b}{Berger et al.\ (2001)}
\tablenotetext{c}{Fynbo et al.\ (2001)}
\tablenotetext{d}{Jha et al.\ 2001; Fruchter et al.\ 2001}
\end{deluxetable}

\begin{deluxetable} {lllll}
\tablewidth{470pt}
\tablecaption{KS test comparison results between models and observed offsets\tablenotemark{a}}
\tablehead{Galaxy mass: & $M_{MW}$ &  $0.1 M_{MW}$  &  $0.01 M_{MW}$   &
$0.001 M_{MW}$  }
\startdata

{\bf WD-NS mergers}      \\ \hline
Standard model & $2.42 \times 10^{-4}$& $3.50 \times 10^{-2}$& $5.87 \times 10^{-1}$& $9.48 \times 10^{-1}$  \\
maximal model\tablenotemark{b}   & $2.94 \times 10^{-4}$\ (B7)& $4.41 \times 10^{-2}$\ (B8)& $5.29 \times
10^{-1}$\ (L1)& $3.26 \times 10^{-1}$\ (F2)  \\
minimal model\tablenotemark{b}  & $1.20 \times 10^{-4}$\ (F2)& $2.53 \times 10^{-2}$\ (F2)& $8.04 \times
10^{-2}$\ (B12)& $6.54 \times 10^{-4}$\ (B12)  \\ [0.5cm]

\multicolumn{5}{l}{\bf WD-BH mergers}      \\ \hline
Standard model & $2.24 \times 10^{-4}$& $3.49 \times 10^{-2}$& $5.87 \times 10^{-1}$& $9.48 \times 10^{-1}$  \\
maximal model  & $2.86 \times 10^{-4}$\ (B6)& $3.92 \times 10^{-2}$\ (B2)& $6.85 \times
10^{-1}$\ (E2)& $9.57 \times 10^{-1}$\ (J) \\
minimal model  & $1.01 \times 10^{-4}$\ (E3)& $1.80 \times 10^{-2}$\ (E3)& $8.23 \times
10^{-2}$\ (G2)& $6.65 \times 10^{-4}$\ (G2)  \\ [0.5cm]

\multicolumn{5}{l}{\bf He-NS mergers}      \\ \hline
Standard model & $2.70 \times 10^{-4}$& $3.61 \times 10^{-2}$& $8.27 \times 10^{-2}$& $6.71 \times 10^{-4}$  \\
maximal model  & $3.57 \times 10^{-4}$\ (B7)& $4.39 \times 10^{-2}$\ (C)& $8.32 \times
10^{-2}$\ (J)& $7.12 \times 10^{-4}$\ (F1)  \\
minimal model  & $2.22 \times 10^{-4}$\ (E3)& $3.57 \times 10^{-2}$\ (F1)& $6.95 \times
10^{-2}$\ (B6)& $6.21 \times 10^{-4}$\ (B1)  \\ [0.5cm]

\multicolumn{5}{l}{\bf He-BH mergers}      \\ \hline
Standard model & $2.84 \times 10^{-4}$& $3.81 \times 10^{-2}$& $8.15 \times 10^{-2}$& $7.07 \times 10^{-4}$  \\
maximal model  & $3.21 \times 10^{-4}$\ (F1)& $3.99 \times 10^{-2}$\ (G2)& $1.00 \times
10^{-1}$\ (I)& $7.98 \times 10^{-4}$\ (L2)  \\
minimal model  & $1.71 \times 10^{-4}$\ (B1)& $3.15 \times 10^{-2}$\ (B4)& $7.87 \times
10^{-2}$\ (B7)& $6.28 \times 10^{-4}$\ (B1)  \\ [0.5cm]

\multicolumn{5}{l}{\bf NS-NS mergers}      \\ \hline
Standard model & $1.84 \times 10^{-4}$& $2.90 \times 10^{-2}$& $3.39 \times 10^{-1}$& $1.04 \times 10^{-2}$ \\
maximal model  & $2.44 \times 10^{-4}$\ (B2)& $3.63 \times 10^{-2}$\ (B1)& $5.72 \times
10^{-1}$\ (M2)& $2.86 \times 10^{-1}$\ (F2)  \\
minimal model  & $1.20 \times 10^{-4}$\ (F2)& $2.14 \times 10^{-2}$\ (F2)& $1.62 \times
10^{-1}$\ (B1)& $2.61 \times 10^{-3}$\ (B1)  \\ [0.5cm]

\multicolumn{5}{l}{\bf NS-BH mergers}      \\ \hline
Standard model & $3.13 \times 10^{-5}$& $4.69 \times 10^{-4}$& $1.25 \times 10^{-6}$& $8.44 \times 10^{-8}$  \\
maximal model  & $2.72 \times 10^{-4}$\ (D2)& $3.16 \times 10^{-2}$\ (D2)& $4.07 \times
10^{-1}$\ (D1)& $2.19 \times 10^{-2}$\ (D1)  \\
minimal model  & $8.37 \times 10^{-6}$\ (O)& $1.80 \times 10^{-6}$\ (O)& $9.78 \times
10^{-10}$\ (O)& $1.09 \times 10^{-10}$\ (O)  \\ [0.5cm]

\enddata
\label{KS}
\tablenotetext{a}{We list the probabilities that the observed
offsets distribution has been drawn from the theoretical one}
\tablenotetext{b}{corresponding models are given in parenthesis}
\end{deluxetable}

\begin{references}

\reference{} Abt, H.\ A.\ 1983, ARA\&A, 21, 343

\reference{} Andersen, M.\ I.\ et al.\ 2000, \aap, 364, L54

\reference{Bel99} Belczynski, K., \& Bulik, T.\ 1999, \aap, 346, 91

\reference{} Belczynski, K., Bulik, T, \& Zbijewski, W.\ 2000,
\aap, 355, 479

\reference{BK01} Belczynski, K., \& Kalogera, V.\ 2001, \apj, 550, L183

\reference{DNS} Belczynski, K., Bulik, T, \& Kalogera, V.\ 2001a,
\apj, submitted [astro-ph/0111452]

\reference{LIGO} Belczynski, K., Kalogera, V., \& Bulik, T.\ 2001b,
in preparation 

\reference{BB98} Berger, E., Kulkarni, S.\ R., \& Frail, D.\ A.\ 2001,
ApJ, submitted [astro-ph/0105081]

\reference{BB98} Bethe, H., \& Brown, G.\ E.\ 1998, \apj , 506, 780

\reference{BB98} Bhattacharya, D., \& van den Heuvel, E.\ P.\ J.\ 1991,
Phys.\ Rep., 203, 1

\reference{BB98} Blaes, O., \& Rajagopal, M.\ 1991, ApJ, 381, 210

\reference{BKD00} Bloom, J.S., Kulkarni, S.R., \& Djorgovski, S.G.\
2001, \apj , accepted [astro-ph/0010176]

\reference{BKD00} Bloom, J.\ S., Sigurdsson, S., \&  Pols, O.\ R.\ 1999,
MNRAS, 305, 763

\reference{BKD00} Bottcher, M., \& Fryer, C.\ L.\ 2001, ApJ, 547, 338

\reference{Brown95} Brown, G.E.\ 1995, \apj , 440, 270

\reference{Brown95} Brown, G.\ E., Lee, C.\ H., Wijers, R.\ A.\ M.\ J.,
Lee, H.\ K., Israelian, G., \& Bethe, H.\ A.\ 2000, New Astronomy,
5, 191

\reference{Brown95} Bulik, T., Belczynski, K.,  \& Zbijewski, W.\ 1999,
MNRAS, 309, 629

\reference{Cappe99} Cappellaro, E., Evans, R., \& Turatto, M.\ 1999,
\aap, 351, 459

\reference{Corde97} Chevalier, R.\ A.\ 2000, in ` Gamma-ray Bursts, 5th Huntsville
Symposium'', eds., R.\ M.\ Kippen et al., AIP Conference Series, 526, 608

\reference{Corde97} Cordes, J., \& Chernoff, D.F.\ 1998, ApJ, 505, 315

\reference{Corde97} Costa E., et al.\ 1997, IAU Circ. 6576, 1

\reference{} Dewi, J.\ D.\ M., \& Tauris, T.\  M.\ 2000,
A\&A, 360, 1043

\reference{Corde97} Duquennoy, A., \& Mayor, M.\ 1991, A\&A,
248, 485

\reference{} Fruchter, A., Burud, I., Rhoads, J., \& Levan, A.\ 2001,
GCN GRB Obesrvation Report No.1087 [http://gcn.gsfc.nasa.gov/gcn/gcn3]

\reference{} Fruchter, A.\ et al.\ 1999, \apj, 519, L13

\reference{} Fryer, C.\ L., Holz, D.\ E., \& Hughes, S.\ A.\ 2001,
ApJ, accepted [astro-ph/0106113]

\reference{} Fryer, C.\ L., \& Woosley, S.\ E.\ 1998, ApJ, 502, L9

\reference{FWH99} Fryer, C.\ L., Woosley, S.\ E., \& Hartmann, D.\ H.\
1999a, \apj , 526, 152

\reference{FWH99} Fryer, C.\ L., Woosley, S.\ E., Herant, M., \& Davies,
M.\ B.\ 1999b, ApJ, 520, 650

\reference{Fyn01} Fynbo, J.\ P.\ U., et al.\ 2001, GCN GRB Obesrvation
Report No.871 [http://gcn.gsfc.nasa.gov/gcn/gcn3]

\reference{G01} Gilmore, G.\ 2001, to appear in `Galaxy Disks and Disk
Galaxies'', eds.\ J.G.\ Funes \& E.M.\ Corsini (San Francisco: ASP)

\reference{G01}  Graziani, C., Lamb, D.\ Q., \& Marion, G.\ H.\ 1999,
A\&AS, 138, 469

\reference{G01} Groot P.\ J., et al.\ 1997a, IAU Circ. 6588, 1

\reference{G01} Groot P.\ J., et al.\ 1997b, IAU Circ. 6584, 1

\reference{} Hamann, W.\ R., Koesterke, L., \&
Wessolowski, U.\ 1995, A\&A, 299, 151

\reference{} Harrison, F.\ A., et al.\ 1999, \apj, 523, L121

\reference{} Hartman, J.\ W.\ 1997, A\&A, 322, 127

\reference{} Holland, S. \ 2001, to appear in the Proceedings of the 
20th Texas Symposium on Relativistic Astrophysics, [astro-ph/0102413]

\reference{Hurle00} Hurley, J.\ R., Pols, O.\ R., \& Tout, C.\ A.\ 
2000, \mnras, 315, 543

\reference{} Jha, S.\ et al.\ 2001, \apj, 554, L155 

\reference{} Kalogera, V.\ 1996, ApJ, 471, 352

\reference{KB96} Kalogera, V., \& Baym, G.A.\ 1996, \apj , 470, L61

\reference{KB96} Kluzniak, W., \& Lee, W.\ H.\ 1998, ApJ, 494, L53

\reference{} Kudritzki, R.\ P., \& Reimers, D.\ 1978, A\&A, 70, 227

\reference{} Kudritzki, R.\ P., Pauldarch, A., Puls, J., \& Abbot,
D.\ C.\ 1989, A\&A, 219, 205

\reference{}  Kuulkers, E., et al.\ 2000, \apj, 538, 638

\reference{} Lee, W.\ H., \& Kluzniak, W.\ 1995, Acta Astronomica,
45, 705

\reference{Lip07} Lipunov, V.\ M., Postnov, K.\ A., \& Prokhorov, M.\ E.\ 
1997, \mnras, 288, 245 

\reference{} Lipunov, V.\ M., Postnov, K.\ A., Prokhorov, M.\ E., \& 
Panchenko, I.\ E.\ 1995, \apj, 454, 593

\reference{} MacFadyen, A., \& Woosley, S. E. 1999, ApJ, 524, 262

\reference{} Madau, P., Ferguson , H.\ C., Dickinson, M.\ E., Giavalisco,
M., Steidel, C.\ C., \& Fruchter, A.\ 1996, \mnras, 283, 1388   

\reference{} Meszaros, P.\ 2000, Nuclear Phys.\ B, 80, 63

\reference{} Meszaros, P., \& Rees, M.\ J.\ 1997, ApJ, 476, 232

\reference{Meu89} Meurs, E.\ J.\ A., \& van den Heuvel, E.\ P.\ J.\ 1989,
\aap, 226, 88

\reference{} Miyamoto, M., \& Nagai, R.\ 1975, \pasj, 27, 533

\reference{} Nieuwenhuijzen, H., \& de Jager, C.\ 1990, A\&A, 231,
134

\reference{} Ostlin, G., Amram, P., Bergvall, N., Masegosa, J., Boulesteix,
J., \& Marquez, I.\ 2001, \aap, 374, 800

\reference{} Paciesas, W. et al., 1999, ApJS, 122, 465

\reference{Pacz90} Paczynski, B.\ 1990, \apj, 348, 485

\reference{} Paczynski, B.\ 1998, \apj, 494, L45

\reference{} Paczynski, B.\ 1999, to appear in `The Largest Explosions
Since the Big Bang: Supernovae and Gamma Ray Bursts", eds., M.\ Livio,
K.\ Sahu, \& N.\ Panagia, Cambridge,  Cambridge University Press
[astro-ph/9909048]

\reference{} Panaitescu, A., \& Kumar, P.\ 2001, \apj, 554, 667


\reference{PB01} Perna, R., \& Belczynski, K.\ 2001, \apj, submitted

\reference{Podsi92} Podsiadlowski, P., Joss, P.C., \& Hsu, J.J.L.\ 1992,
\apj , 391, 246

\reference{} Portegies-Zwart, S.\ F., \& Yungelson, L.\ R.\ 1998,
A\&A, 332, 173

\reference{} Press, W.\ H., Teukolsky, S.\ A., Vetterling, W.\ T., \&
Flannery, B.\ P.\ 1992, ``Numerical Recipes in C'', Second Edition,
Cambridge University Press

\reference{} Rowan-Robinson, M.\ 1999, Ap\&SS, 266, 291

\reference{} Ruffert, M., Janka, , H., Takahashi, K., \& Schaefer,
G.\ 1997, A\&A, 319, 122

\reference{} Scalo, J.\ M.\ 1986, Fundam. Cosmic Phys., 11, 1

\reference{}  Stanek, K.\ Z., Garnavich, P.\ M., Kaluzny, J., Pych, W., \&
Thompson, I.\ 1999, \apj, 522, L39

\reference{} Vassiliadis, E., \& Wood, P.\ R.\ 1993, ApJ, 413, 641

\reference{}  Webbink, R.\ F.\ 1984, ApJ, 277, 355

\reference{Woosl86} Woosley, S.E.\ 1986, in ``Nucleosynthesis and Chemical
Evolution'', 16th Saas-Fee Course, eds. B. Hauck et al., Geneva Obs., 
p.\ 1

\reference{} Woosley, S.E.\ 1993, \apj, 405, 273

\reference{} Woosley, S.\ E.\ 2000, in ``Gamma-ray Bursts, 5th Huntsville
Symposium'', eds., R.\ M.\ Kippen et al., AIP Conference Series, 526, 555

\reference{} Zhang,  W., \& Fryer, C.\ L.\ 2001, ApJ, 550, 357



\end{references}
\end{document}